\documentclass[%
twocolumn,
preprintnumbers,
 amsmath,amssymb,
prc,
]{revtex4-2}

\usepackage{graphicx}
\usepackage{dcolumn}
\usepackage{bm}

\usepackage{aas_macros}
\usepackage{color}
\usepackage{amsmath}

\usepackage{setspace}
\usepackage{subfigure}

\newcommand{\average}[1]{\ensuremath{\langle#1\rangle}}

\usepackage{hyperref}
\hypersetup{
    colorlinks=true,
    linkcolor=blue,
    citecolor=blue,        
    filecolor=magenta,      
    urlcolor=blue,
}

\usepackage{braket}

\begin{document}

\preprint{LA-UR-25-27311}
\title{
Modeling direct and pre-equilibrium processes of neutron-induced reactions with noniterative finite amplitude method and distorted--wave Born approximation
}



\author{Hirokazu Sasaki}
\email{hsasaki@lanl.gov}

\affiliation{Theoretical Division, Los Alamos National Laboratory, Los Alamos, New Mexico 87545, USA}
\author{Toshihiko Kawano}

\affiliation{Theoretical Division, Los Alamos National Laboratory, Los Alamos, New Mexico 87545, USA}

\author{Marc Dupuis}
\affiliation{CEA, DAM, DIF, F-91297 Arpajon, France}
\affiliation{Universit\'{e} Paris-Saclay, CEA, Laboratoire Mati\`{e}re sous Conditions Extr\^{e}mes, 91680 Bruy\`{e}res-Le-Ch\^{a}tel, France}




\date{\today}

\begin{abstract}


We develop a calculation method for describing the direct and pre-equilibrium processes in neutron-induced reactions based on the framework of noniterative finite amplitude method (FAM) and distorted-wave Born approximation (DWBA). The noniterative FAM is used to derive equations of quasiparticle random-phase approximation (QRPA) for neutron-induced inelastic scatterings to both the discrete and continuum states in a consistent manner. The Skyrme force is employed as an interaction between the projectile neutron with nucleons inside the target nucleus. We apply this method to the neutron-induced reaction on $^{208}$Pb. We demonstrate that the calculated differential inelastic scattering cross sections to low-lying states reproduce available experimental data without any phenomenological parameters that are often introduced in conventional DWBA calculations. The calculated double differential cross section to the continuum state also agrees with the experimental data in the energy region relevant to the direct and pre-equilibrium processes. These results are used to investigate the spin distribution of the populated states in the residual nucleus.

\end{abstract}

\maketitle


\section{Introduction}
Modeling neutron-induced reactions based on a developed theoretical framework is essential for nuclear astrophysics \cite{Burbidge:1957vc,ARNOULD200797,Mumpower:2015ova,Kajino:2019abv} and for practical applications \cite{Hayes:2020cji,NEUDECKER2021108345}. At incident neutron energies above 10 MeV, 
the neutron--induced reactions on medium-
to heavy-mass nuclei proceed through three main mechanisms: compound nucleus formation, direct reaction, and pre-equilibrium reaction \cite{gadioli1992pre}. The compound nucleus formation occurs through multiple scatterings of incident neutrons with nucleons inside the target nucleus, eventually leading to thermal equilibrium in the residual system. The decay of the compound nucleus is described by the statistical Hauser-Feshbach theory \cite{Hauser:1952zz}, which provides the low-energy part of  the emitted neutron spectrum. The direct reaction occurs when the incident neutron interacts with a small number of nucleons inside the target nucleus on a fast time scale. The differential cross section of the direct process is calculated from the optical model, distorted-wave Born approximation (DWBA), and Coupled-Channels (CC) methods \cite{glendenning2004direct}. The contribution from the direct process to the outgoing neutron spectrum is confirmed in a few peaks in the high energy region.\\
\indent The pre-equilibrium reaction is an intermediate process between the compound nucleus and direct reaction mechanism. Its contribution to the neutron energy spectrum appears as a smooth tail extending between the low-energy evaporative peak and the high-energy discrete peaks associated with direct reactions. Theoretical approaches to modeling pre-equilibrium reactions are broadly classified into classical/semiclassical and quantum mechanical models. Among the classical/semiclassical models, the exciton model \cite{GriffinPRL1966} and the hybrid model \cite{BlannPRL1971} have been widely used to describe the neutron energy spectrum. These models have been extended to account for the angular distribution of emitted neutrons \cite{mantzouranis1976generalized,BlannPRC1984}. Some efforts have been made to compensate deficiencies inherent in the classical approaches in the past, e.g., estimation of the transferred angular momentum by a microscopic model \cite{KAWANO2006774}, and the collective enhancement effects in the residual nucleus \cite{MumpowerPRC2023}.
\\
\indent Quantum mechanical models of 
pre-equilibrium reactions avoid the phenomenological assumptions required in classical/semiclassical models, which allows for a more general and microscopic treatment. The multistep compound (MSC) and multistep direct (MSD) processes were originally proposed by Feshbach, Kerman, and Koonin (FKK) \cite{FESHBACH1980429}. Various MSD models have been developed under different statistical assumptions \cite{Tamura-Udagawa-Lenske,NISHIOKA1989195,KONING1991216,KawanoPRC2002}. The MSD process becomes dominant, and the one-step contribution is larger than the two-step contribution at incident neutron energies as low as 14 MeV \cite{ChadwickPRC1993}. In the one-step process, the incident neutron induces 1p-1h excitations within the target nucleus. These excitations can be microscopically calculated using the random-phase approximation (RPA) \cite{ring2004nuclear}. This reaction mechanism is modeled by combining RPA with the DWBA framework \cite{Tamura-Udagawa-Lenske,Dupuis:2011zza}. The pre-equilibrium one-step process and the direct reaction have been treated consistently without artificial separation through a self-consistent RPA   incorporating Gogny D1S force and the density-dependent M3Y forces to describe the two-body interaction between nucleons \cite{Dupuis:2011zza}. This microscopic approach has been further extended to axially deformed nuclei, where the scattering potential is computed using the Jeukenne-Lejeune-Mahaux (JLM) folding model, combined with the transition densities obtained from quasiparticle RPA (QRPA) calculations \cite{Dupuis:2015uqy,dupuis2017microscopic,Dupuis2019}.\\
\indent The finite amplitude method (FAM) \cite{Nakatsukasa:2007qj,Avogadro:2011gd} is an efficient technique for solving the fully self-consistent (Q)RPA equations. FAM has been successfully applied to a wide range of studies, including multipole collective excitations \cite{Inakura:2009vs,Stoitsov:2011zz,Hinohara:2013qda,Kortelainen:2015gxa,Oishi:2015lph,Sasaki:2022ipn,SasakiPRC2023,Washiyama:2023jxg,Frosini:2023bym,Li:2024ivd}, weak interactions \cite{Mustonen:2014bya,Shafer:2016etk,Ney:2020mnx,Hinohara:2022uip,Liu:2023xlv}, the fission calculation \cite{Washiyama:2020qfr}, and the nuclear level density \cite{BJELCIC2025109387}. However, FAM-QRPA has not yet been applied to neutron-induced reactions. In such reactions, various natural and unnatural parity transitions of the residual nucleus must be taken into account. 



\indent In this paper we develop a calculation method for the direct and one-step pre-equilibrium processes in neutron-induced reactions based on FAM-QRPA combined with DWBA. This study extends our previous work on non-iterative FAM \cite{Sasaki:2022ipn,SasakiPRC2023}  by replacing the electromagnetic transition operator with the nuclear interaction between the incident neutron and nucleons in the target nucleus. For the nuclear interaction, we adopt the Skyrme effective interaction, which is widely used in nuclear structure calculations. Within this unified theoretical framework, we calculate both differential cross sections for discrete final states and double-differential cross sections for continuum states. Simultaneously, we provide a prediction of the spin distribution of the residual nucleus.

\section{Theory}

\label{sec:theory}

\subsection{QRPA equations derived from FAM}
We employ a noniterative finite amplitude \cite{Sasaki:2022ipn,SasakiPRC2023} to derive the QRPA equation for neutron--induced inelastic scattering. In the general framework of FAM \cite{Nakatsukasa:2007qj,Avogadro:2011gd}, the residual interaction is numerically expressed with small parameter $\eta$. Then the forward and backward amplitudes of a frequency $\omega$ are calculated by solving iteratively a linear response equation of the Time-Dependent Hartree-Fock (TDHF) equation with the residual interaction and the external field $F$ inducing collective motion of nucleons inside the target nucleus. The obtained amplitudes are applicable to the transition strength and cross section of the giant resonance. The noniterative FAM \cite{Sasaki:2022ipn,SasakiPRC2023} avoids the iterative procedure as in the general FAM with the explicit linearlization of the residual interaction in the limit of $\eta\to0$. The QRPA equation is obtained from the linear response equation of the TDHF equation under this explicit linearization. The QRPA equation for even-even nuclei has been derived in Ref.~\cite{SasakiPRC2023} based on the framework of the Hartree-Fock+Bardeen–Cooper–Schrieffer (HF+BCS) calculation,
\begin{equation}
\left(
\begin{array}{cc}
     A -\omega& B \\
     B^{*} & A^{*}+\omega 
\end{array}
\right)_{\mu\nu q,\alpha\beta q^{\prime}}
\left(
\begin{array}{c}
     X_{\alpha\beta}^{q^{\prime}}\\
     Y_{\alpha\beta}^{q^{\prime}}
\end{array}
\right)
=-\zeta^{\tau}_{\mu\nu}
\left(
\begin{array}{c}
     f_{\mu\nu}^{q}\\
     f_{\nu\mu}^{q}
\end{array}
\right),
\label{eq:QRPAeq}
\end{equation}
where $A$ and $B$ are QRPA matrices, and $X_{\alpha\beta}^{q^{\prime}}(Y_{\alpha\beta}^{q^{\prime}})$ is the forward (backward) amplitude of nucleon $q^{\prime}$. The indices $\alpha,\beta,\nu,\mu$ label the single-particle states of the HF equation, and the configuration space of Eq.~(\ref{eq:QRPAeq}) is restricted to $\mu\geq\nu,$ $\alpha\geq\beta$. The factor $\zeta^{\tau}_{\mu\nu}\equiv u_{\mu}v_{\nu}+\tau u_{\nu}v_{\mu}$ is composed of the BCS parameters  $u_{\mu},v_{\mu}\geq0$ from the HF+BCS calculation and the index $\tau=\pm1$ relevant to the time-reversal invariance for the external field $F$. The components $f_{\mu\nu}^{q}$ and $f_{\nu\mu}^{q}$ in the right hand side of Eq.~(\ref{eq:QRPAeq}) are described by
\begin{equation}
\label{eq:external field f}
f_{\mu\nu}^{q}=\int\mathrm{d}^{3}r\ \phi_{\mu}^{q*}F\phi_{\nu}^{q},\ \
f_{\nu\mu}^{q}=\int\mathrm{d}^{3}r\ \phi_{\nu}^{q*}F\phi_{\mu}^{q},
\end{equation}
where $\phi^{q}_{\mu(\nu)}\equiv\phi_{\mu(\nu)}(\vec{r},\sigma,q)$ is a single-particle state of the HF calculation described in the spaces of the coordinate $\vec{r}$, the spin $\sigma$, and the isospin $q=(n,p)$ \cite{Vautherin:1973zz}. The derivation in Ref.~\cite{SasakiPRC2023} assumed the Hermitian one-body external field $F$ for the calculation of the M1 transition. However Eq.~(\ref{eq:QRPAeq}) is still applicable to the non-Hermitian operator $F$ of the neutron-induced inelastic scattering which satisfies $TFT^{-1}=\tau F^{\dagger}$ with $\tau=\pm1$ where $T$ is the time reversal operator. The QRPA matrices $A$ and $B$ are determined without any information of the external field, and their matrix components are calculated from the same procedure as done in Ref.~\cite{SasakiPRC2023}. We calculate the inelastic scattering operator $F$ and apply it to Eq.~(\ref{eq:QRPAeq}). Then the obtained amplitudes are used for the calculation of  the inelastic scattering differential cross section.

\subsection{FAM-QRPA+DWBA calculations}
The external field for the inelastic scattering $V_{\mathrm{inl}}$ applicable to Eq.~(\ref{eq:external field f}) should include information of the wave functions of both incoming and outgoing neutrons. We employ the framework of DWBA and regard the wave functions of incoming and outgoing neutrons as distorted waves of the elastic scatterings that are calculated from the optical model. The operator is described by
\begin{equation}
\label{eq:external dwba}
    V_{\mathrm{inl}}=\int\mathrm{d}^{3}r_{2}\chi^{(-)*}_{\alpha}v_{12}(r_{12})(1-P_{r}P_{\sigma}P_{\tau})\chi^{(+)}_{\beta},
\end{equation}
where $r_{12}=|\vec{r}_1-\vec{r}_2|$, $\chi^{(-)}_{\alpha}\equiv\chi^{(-)}_{\alpha}(\vec{r}_{2},\vec{k}_{\alpha},\Sigma_{\alpha})$ and $\chi^{(+)}_{\beta}\equiv\chi^{(+)}_{\beta}(\vec{r}_{2},\vec{k}_{\beta},\Sigma_{\beta})$ are distorted waves of outgoing and incoming neutrons whose momentum are $\vec{k}_{\alpha}$ and $\vec{k}_{\beta}$. $\Sigma_{\alpha(\beta)}(=\pm1/2)$ is the $z$-component in the neutron spin space. $v_{12}$ is the nuclear force between the nucleon inside the target nucleus (index 1) and the incident/scattered neutron (index 2). $P_{r}, P_{\sigma},$ and $P_{\tau}$ are the exchange operators between the indices $1$ and $2$ in the coordinate space, the spin space, and the isospin space. 
In general, the nuclear force $v_{12}$ in Eq.~(\ref{eq:external dwba}) includes various natural parity ($0^{+},1^{-},2^{+}$, $\ldots$) and unnatural parity ($1^{+},2^{-},3^{+}$, $\ldots$) multipole transitions. All of these transitions contribute to the double differential cross section \cite{Dupuis:2011zza}. To calculate the total contribution from the various $J^{\Pi}\ (J=0,1,2,\ldots, \Pi=\pm1)$ transitions, we decompose Eq.~(\ref{eq:external dwba}) by using the multipole operators $\{F^{M}_{\mathrm{inl}}(J^{\Pi})\}_{M=-J,..,J}$ inducing the $J^{\Pi}$ transition,
\begin{equation}
\label{eq:expansion external}
    V_{\mathrm{inl}}=\sum_{JM\Pi}F^{M}_{\mathrm{inl}}(J^{\Pi}).
\end{equation}
Each operator in the right hand side of Eq.~(\ref{eq:expansion external}) satisfies the time-reversal invariance, $TF^{M}_{\mathrm{inl}}(J^{\Pi})T^{-1}=\tau F^{M}_{\mathrm{inl}}(J^{\Pi})^{\dagger}$ with $\tau=+1 (-1)$ for the natural parity (unnatural parity) transition. Therefore, the QRPA equation in Eq.~(\ref{eq:QRPAeq}) is applicable for the each operator of $F^{M}_{\mathrm{inl}}(J^{\Pi})$ and the double differential cross section of neutron-induced inelastic scattering for the even-even target nucleus is derived from Eqs.~(\ref{eq:QRPAeq})-(\ref{eq:expansion external}),
\begin{equation}
\label{eq:double differential cs}
    \frac{\mathrm{d}^{2}\sigma}{\mathrm{d}E_x\mathrm{d}\Omega_{\alpha}}=\frac{\mu_{\alpha}\mu_{\beta}}{(2\pi\hbar^{2})^{2}}\frac{k_{\alpha}}{k_{\beta}}\frac{1}{2}\sum_{\Sigma_\alpha\Sigma_\beta}\sum_{J\Pi M}\frac{\mathrm{d}B(\omega;F^{M}_{\mathrm{inl}}(J^{\Pi}))}{\mathrm{d}E_x},
\end{equation}
\begin{equation}
 \frac{\mathrm{d}B(\omega;F^{M}_{\mathrm{inl}}(J^{\Pi}))}{\mathrm{d}E_x}=-\frac{1}{\pi}\mathrm{Im}\sum_q\sum_{\substack{\mu\nu\in q\\ \mu\geq\nu}}\zeta^{\tau}_{\mu\nu}(f_{\mu\nu}^{q*}X_{\mu\nu}^{q}+f_{\nu\mu}^{q*}Y_{\mu\nu}^{q}),\label{eq:transition strength QRPA+DWBA}
 \end{equation}
 where $\mathrm{d}\Omega_{\alpha}=\mathrm{d}(\cos\theta_{k_\alpha})\mathrm{d}\phi_{k_\alpha}$ is the solid angle of $\vec{k}_{\alpha}$, $\mu_\alpha$ and $\mu_\beta$ are reduced masses
 of final and initial channels, $E_x=\mathrm{Re}(\omega)$ is the excitation energy of the residual nucleus given by $E_{x}=\hbar^2k_\beta^2/2\mu_\beta-\hbar^2k_\alpha^2/2\mu_\alpha$. The factor $1/2$ and the sum $\sum_{\Sigma_\beta}$ in Eq.~(\ref{eq:double differential cs}) represent the average over the initial neutron spin polarization. Equation~(\ref{eq:transition strength QRPA+DWBA}) shows the partial contribution from the $J^{\Pi}$ transition strength. The differential cross section for the excitation to a discrete level $J^{\Pi}$ at $E_x$ is given by
 \begin{equation}
 \label{eq:differential cs discrete}
     \frac{\mathrm{d}\sigma}{\mathrm{d}\Omega_{\alpha}}=\frac{\mu_{\alpha}\mu_{\beta}}{(2\pi\hbar^{2})^{2}}\frac{k_{\alpha}}{k_{\beta}}\frac{1}{2}\sum_{\Sigma_\alpha\Sigma_\beta M}\int_{E_x-\Delta}^{E_x+\Delta}\mathrm{d}E^{\prime}_x\frac{\mathrm{d}B(\omega;F^{M}_{\mathrm{inl}}(J^{\Pi}))}{\mathrm{d}E^{\prime}_x},
 \end{equation}
 where $\Delta$ is the width for the energy integration of the discrete level. The transition strength in Eq.~(\ref{eq:transition strength QRPA+DWBA}) associated with the discrete level is  broadened and smoothed by the Lorentzian width $\gamma=2\mathrm{Im}(\omega)$. The value of $\Delta$ should be larger than that of $\gamma$ for the energy integration in Eq.~(\ref{eq:differential cs discrete}). On the other hand, $\Delta$ should be small to separate the discrete level from other excitations.


 \subsection{Nuclear force}
\label{sec:nuclear force}
In order to calculate the external field in Eq.~(\ref{eq:external dwba}), we need the nuclear force $v_{12}(r_{12})$ between the nucleons inside the target nucleus (index 1) and the projectile or scattered neutron (index 2). We assume that the nuclear force corresponds to the Skyrme force
whose parameters such as $t_i,x_i (i=0,1,2,3), W_0,\alpha$ are determined from nuclear structure \cite{ring2004nuclear,Bender:2003jk},

\begin{eqnarray}
\label{eq:skyrme}
v_{12}&=&t_{0}(1+x_{0}P_{\sigma})\delta(\vec{r}_{1}-\vec{r}_{2}) \nonumber\\
&+&\frac{1}{2}t_{1}(1+x_{1}P_{\sigma})[\vec{k}^{\dagger2}_{12}\delta(\vec{r}_{1}-\vec{r}_{2})+\delta(\vec{r}_{1}-\vec{r}_{2})\vec{k}^{2}_{12}]\nonumber\\
&+&t_{2}(1+x_{2}P_{\sigma})\vec{k}^{\dagger}_{12}\delta(\vec{r}_{1}-\vec{r}_{2})\vec{k}_{12}\nonumber\\
&+&\frac{1}{6}t_{3}(1+x_{3}P_{\sigma})\delta(\vec{r}_{1}-\vec{r}_{2})\rho^{\alpha}(\frac{\vec{r}_{1}+\vec{r}_{2}}{2})\nonumber\\
&+&iW_{0}(\sigma_{1}+\sigma_{2})\vec{k}^{\dagger}_{12}\delta(\vec{r}_{1}-\vec{r}_{2})\vec{k}_{12},
\end{eqnarray}
where $\vec{k}_{12}=(\nabla_{1}-\nabla_{2})/2i$, $P_{\sigma}=(1+\sigma_{1}\sigma_{2})/2$, and $\rho$ is the total nucleon density. The Skyrme force is a local interaction proportional to $\delta(\vec{r}_{1}-\vec{r}_{2})$. Then this interaction satisfies the zero-range approximation \cite{SATCHLER19641}, and the integration in Eq.~(\ref{eq:external dwba}) becomes an one-body operator including $\vec{r}_1,\nabla_1,$ and $\sigma_1$. The exchange operator of the isospin $P_{\tau}$ in Eq.~(\ref{eq:external dwba}) is written as $\delta_{qn}$ where the $q$ is the isospin of the nucleon inside the target nucleus. \\
\indent The calculation of $f_{\mu\nu}^{q}$ in Eq.~(\ref{eq:external field f}) is straightforward by using Eqs.~(\ref{eq:external dwba}) and (\ref{eq:skyrme}). For example, the first line on the right hand side of  Eq.~(\ref{eq:skyrme}) is symmetric towards the exchange between $\vec{r}_{1}$ and $\vec{r}_{2}$, which results in $P_{r}=1$ for this term. Then the contribution to $f_{\mu\nu}^{q}$ in Eq.~(\ref{eq:external field f}) is written as
\begin{eqnarray}
\label{eq:t0 contribution}
&&\int\mathrm{d}^{3}r{d}^{3}r_{2}\phi_{\mu}^{q*}\chi^{(-)*}_{\alpha}t_{0}(1+x_{0}P_{\sigma})
\delta(\vec{r}-\vec{r}_{2})(1-P_{r}P_{\sigma}P_{\tau})\chi^{(+)}_{\beta}\phi_{\nu}^{q}\nonumber\\
&=&\int\mathrm{d}^{3}r\phi^{q*}_{\mu}\chi^{(-)*}_{\alpha}t_{0}(1+x_{0}P_{\sigma})(1-P_{\sigma}\delta_{nq})\chi^{(+)}_{\beta}\phi^{q}_{\nu} \nonumber\\
&=&(b_{0}-b_{0}^{\prime}\delta_{nq})\int\mathrm{d}^{3}r\ \chi^{(-)*}_{\alpha}\chi_{\beta}^{(+)}\phi^{q*}_{\mu}\phi_{\nu}^{q} \nonumber\\
&+&(\Tilde{b}_{0}-\Tilde{b}_{0}^{\prime}\delta_{nq})\int\mathrm{d}^{3}r\ (\chi^{(-)*}_{\alpha}\vec{\sigma}\chi_{\beta}^{(+)})\cdot(\phi^{q*}_{\mu}\vec{\sigma}\phi_{\nu}^{q}),
\end{eqnarray}
where $\vec{r}=\vec{r}_{1}$ and $P_{\sigma}^2=1$. In the third and forth lines, we use
$b$ parameters of the Skyrme force,
$b_{0}=t_{0}(1+\frac{x_{0}}{2}), b_{0}^{\prime}~=t_{0}(\frac{1}{2}+x_{0}), \Tilde{b}_{0}=\frac{t_{0}x_{0}}{2},$ and $\Tilde{b}_{0}^{\prime}=\frac{t_{0}}{2}$ given in Refs.~\cite{Maruhn:2013mpa,Vesely:2009eb}. The contributions to $f_{\mu\nu}^{q}$ from other terms in Eq.~(\ref{eq:skyrme}) are calculated with the Skyrme $b$ parameters as in Eq.~(\ref{eq:t0 contribution}). The sign of the exchange operator $P_{r}=\pm1$ is negative and positive for the Skyrme forces including $\vec{k}_{12}$ and $\vec{k}_{12}^{2}$ in Eq.~(\ref{eq:skyrme}) due to $\vec{k}_{21}=-\vec{k}_{12}$. As done in Ref.~\cite{Sasaki:2022ipn}, we ignore the tensor terms which maybe proportional to $\Tilde{b}_{1}$ or $\Tilde{b}_{1}^{\prime}$ from the second and third lines of Eq.~(\ref{eq:skyrme}). Then, the $f_{\mu\nu}^{q}$ in Eq.~(\ref{eq:external field f}) is given by
\begin{widetext}
\begin{equation}
\begin{split}
\label{eq:inelastic f QRPA}
f_{\mu\nu}^{q}&=\left(
b_{0}-\delta_{nq}b_{0}^{\prime}
\right)\int\mathrm{d}^{3}r\ 
\phi_{\mu}^{q*}\phi_{\nu}^{q}\chi_{\alpha}^{(-)*}\chi_{\beta}^{(+)}\\
&+\left(
b_{1}-\delta_{nq}b_{1}^{\prime}
\right)\int\mathrm{d}^{3}r\ \left(
\phi_{\mu}^{q*}\phi_{\nu}^{q}\nabla\chi_{\alpha}^{(-)*}\cdot\nabla\chi_{\beta}^{(+)}
+\chi_{\alpha}^{(-)*}\chi_{\beta}^{(+)}\nabla\phi_{\mu}^{q*}\cdot\nabla\phi_{\nu}^{q}\right)\\
&-\left(
b_{2}-\delta_{nq}b_{2}^{\prime}
\right)\int\mathrm{d}^{3}r\ 
\nabla^{2}(\phi_{\mu}^{q*}\phi_{\nu}^{q})
\chi_{\alpha}^{(-)*}\chi_{\beta}^{(+)}\\
&+(b_{3}-\delta_{nq}b_{3}^{\prime})\int\mathrm{d}^{3}r\ \frac{(\alpha+2)(\alpha+1)}{2}\frac{2}{3}(\rho_{0})^{\alpha}
\phi_{\mu}^{q*}\phi_{\nu}^{q}\chi_{\alpha}^{(-)*}\chi_{\beta}^{(+)}\\
&-\left(
b_{4}+\delta_{nq}b_{4}^{\prime}
\right)\int\mathrm{d}^{3}r\ \left\{ 
\phi_{\mu}^{q*}\phi_{\nu}^{q}\nabla\chi_{\alpha}^{(-)*}\cdot\left(
    -i
    \right)\left(
    \nabla\times\vec{\sigma}
    \right)\chi_{\beta}^{(+)}
+\chi_{\alpha}^{(-)*}\chi_{\beta}^{(+)}\nabla\phi_{\mu}^{q*}\cdot\left(
    -i
    \right)\left(
    \nabla\times\vec{\sigma}
    \right)\phi_{\nu}^{q}\right\}\\
    \\
&-2(b_{1}-\delta_{nq}b_{1}^{\prime})\int\mathrm{d}^{3}r\ \left\{
\frac{1}{2i}(\phi_{\mu}^{q*}\nabla\phi_{\nu}^{q}-\phi_{\nu}^{q}\nabla\phi_{\mu}^{q*})\cdot\frac{1}{2i}(\chi_{\alpha}^{(-)*}\nabla\chi_{\beta}^{(+)}-\chi_{\beta}^{(+)}\nabla\chi_{\alpha}^{(-)*})
\right\} \\
&-(b_{4}+\delta_{nq}b_{4}^{\prime})\int\mathrm{d}^{3}r\ \left\{
\frac{1}{2i}(\phi_{\mu}^{q*}\nabla\phi_{\nu}^{q}-\phi_{\nu}^{q}\nabla\phi_{\mu}^{q*})\cdot\nabla\times(\chi_{\alpha}^{(-)*}\vec{\sigma}\chi_{\beta}^{(+)})\right.\\
&\left. +\frac{1}{2i}(\chi_{\alpha}^{(-)*}\nabla\chi_{\beta}^{(+)}-\chi_{\beta}^{(+)}\nabla\chi_{\alpha}^{(-)*})\cdot\nabla\times(\phi_{\mu}^{q*}\vec{\sigma}\phi_{\nu}^{q})
\right\}\\
&+(\tilde{b}_{0}-\delta_{nq}\tilde{b}_{0}^{\prime})\int\mathrm{d}^{3}r\ (\chi_{\alpha}^{(-)*}\vec{\sigma}\chi_{\beta}^{(+)})\cdot(\phi_{\mu}^{q*}\vec{\sigma}\phi_{\nu}^{q})\\
&+(\tilde{b}_{3}-\delta_{nq}\tilde{b}_{3}^{\prime})\int\mathrm{d}^{3}r\ \frac{(\alpha+2)(\alpha+1)}{2}\frac{2}{3}(\rho_{0})^{\alpha}(\chi_{\alpha}^{(-)*}\vec{\sigma}\chi_{\beta}^{(+)})\cdot(\phi_{\mu}^{q*}\vec{\sigma}\phi_{\nu}^{q})\\
&+(\tilde{b}_{2}-\delta_{nq}\tilde{b}_{2}^{\prime})\int\mathrm{d}^{3}r\ \nabla(\chi_{\alpha}^{(-)*}\vec{\sigma}\chi_{\beta}^{(+)})\cdot\nabla(\phi_{\mu}^{q*}\vec{\sigma}\phi_{\nu}^{q}),
\end{split}
\end{equation}
\end{widetext}
where $\rho_{0}=\sum_{qi}\phi^{q*}_{i}\phi^{q}_{i}$ is the density of the nucleon inside the target nucleus. The above equation can be calculated in the similar procedure to obtain RPA matrices by replacing $\phi^{q^{\prime}*}_{j}\phi_{n}^{q^{\prime}}$ with $\chi_{\alpha}^{(-)*}\chi_{\beta}^{(+)}$ in the Eqs.~(32) and (39) of Ref.~\cite{Sasaki:2022ipn}. The derivative and the spin density such as $\chi_{\alpha}^{(-)*}\nabla\chi_{\beta}^{(+)}$ and $\chi_{\alpha}^{(-)*}\vec{\sigma}\chi_{\beta}^{(+)}$ are originated from $\vec{k}_{12}$ and $P_{\sigma}$ in Eq.~(\ref{eq:skyrme}) and the Pauli-blocking factor $(1-P_{r}P_{\sigma}P_{\tau})$. We remark that a factor $(\alpha+2)(\alpha+1)/2$ is multiplied in the three-body term ($\propto t_3$) in Eq.~(\ref{eq:skyrme}) following Refs.~\cite{Davies:1974ell,Sharp1973130}. This factor of the three-body term corrects the Skyrme force usually used in nuclear structure calculations to that of nuclear reaction calculations. Among the various terms in Eq.~(\ref{eq:inelastic f QRPA}), two-body terms and the three-body terms associated with $b_i,b_i^{\prime},\tilde{b}_i,\tilde{b}_i^{\prime}(i=0,3)$ that are proportional to
 $\chi_{\alpha}^{(-)*}\chi_{\beta}^{(+)}$ or $\chi_{\alpha}^{(-)*}\vec{\sigma}\chi_{\beta}^{(+)}$ play the dominant contribution to the strength of the transition \cite{Davies:1974ell}. In order to simplify the calculation, we ignore the terms including the derivatives of the distorted wave functions. This treatment enables the easy derivation of the partial contribution from the multipole transitions in Eq.~(\ref{eq:inelastic f QRPA}) used for Eq.~(\ref{eq:transition strength QRPA+DWBA}).

\subsection{Multipole transitions}
\label{sec:multipole}
From the products $\chi_{\alpha}^{(-)*}\chi_{\beta}^{(+)}$ and $\chi_{\alpha}^{(-)*}\vec{\sigma}\chi_{\beta}^{(+)}$ in Eq.~(\ref{eq:inelastic f QRPA}), we abstract operators for  natural parity ($0^{+},1^{-},2^{+}$, $\ldots$) and unnatural parity ($1^{+},2^{-},3^{+}$, $\ldots$) multipole transitions as in Eq.~(\ref{eq:expansion external}). In the spherical coordinate $(r,\theta,\phi)$, the electric E$L(L=0,1,2,\ \ldots)$ transition is originated from the spherical harmonics $Y_{LM}(\theta,\phi)$ \cite{ring2004nuclear}. On the other hand, the operator for the magnetic M$L(L=1,2,\ \ldots)$ transition is described by a linear combination of $\vec{\sigma}\cdot\nabla(r^{L}Y_{LM})$ and $\vec{l}\cdot\nabla(r^{L}Y_{LM})$ where $\vec{l}$ is the angular momentum operator \cite{ring2004nuclear}. \\
\indent The product $\chi_{\alpha}^{(-)*}\chi_{\beta}^{(+)}$ is a function in the coordinate space $\vec{r}$ without any operator of $\nabla$ and $\vec{\sigma}$. Therefore, this product could be expanded by spherical harmonics only,
\begin{eqnarray}
\label{eq:expansion chiproduct}
    \chi_{\alpha}^{(-)*}\chi_{\beta}^{(+)}&=&\sum_{LM}\lambda_{LM}^{0}(r)Y_{LM}(\theta,\phi),\\
\lambda_{LM}^{0}(r)&=&\int\mathrm{d}\Omega Y_{LM}^{*}(\theta,\phi)\chi_{\alpha}^{(-)*}\chi_{\beta}^{(+)}.\label{eq:lambda0}
\end{eqnarray}
From the expansion of Eq.~(\ref{eq:expansion chiproduct}), the partial contribution from natural parity transitions $F^{M}_{\mathrm{inl}}(J^{\Pi})\ (J^{\Pi}=0^{+},1^{-},2^{+},\ \ldots)$ in Eq.~(\ref{eq:expansion external}) are calculated. For example, the partial contribution of the $J^{\Pi}=L^{(-)^L}\ (L=0,1,...)$ transition in the first line in the right hand side of  Eq.~(\ref{eq:inelastic f QRPA}) is described by
\begin{equation}
\label{eq:EL contribution e.g.}
    \left(
b_{0}-\delta_{nq}b_{0}^{\prime}
\right)\int\mathrm{d}^{3}r\ 
\phi_{\mu}^{q*}\sum_{M}\lambda_{LM}^{0}(r)Y_{LM}\phi_{\nu}^{q}.
\end{equation}
The gradient $\nabla$ in the spherical coordinate $(r,\theta,\phi)$ is described by
\begin{eqnarray}
    \nabla=\vec{e}_{r}\partial_{r}-\frac{i\vec{e}_{r}\times\vec{l}}{r},
\end{eqnarray}
which indicates that the terms proportional to
$\nabla\phi_{\mu}^{q*}\cdot\nabla\phi_{\nu}^{q}$
or $\nabla\chi_{\alpha}^{(-)*}\cdot\nabla\chi_{\beta}^{(+)}$ in Eq.~(\ref{eq:inelastic f QRPA}) only induce natural parity transitions because the product of $\nabla$ does not change both the angular momentum and parity of the residual target nucleus: $\int\mathrm{d}\Omega\nabla Y^{*}_{L^{\prime}M^{\prime}}\cdot\nabla Y_{LM}=-\int\mathrm{d}\Omega\frac{1}{r^{2}}(\vec{l}Y^{*}_{L^{\prime}M^{\prime}})\cdot(\vec{l}Y_{LM})\propto \delta_{L^{\prime}L}$. The Laplacian $\nabla^{2}$ does not induce any unnatural parity transition. Since the product $\vec{\sigma}\cdot\vec{l}$ is a scalar for spatial rotation, the spin--orbit term of Eq.~(\ref{eq:inelastic f QRPA}) only induces natural parity transitions. In short, there is no contribution to the unnatural parity transitions from the first-fifth lines in the right hand side of Eq.~(\ref{eq:inelastic f QRPA}), and the partial contributions of $J^{\Pi}=L^{(-)^L}\ (L=0,1,...)$ transitions are calculated in the same way as Eq.~(\ref{eq:EL contribution e.g.}).

The unnatural parity transitions ($1^{+},2^{-},3^{+}$..) are induced by a vector term $\chi_{\alpha}^{(-)*}\vec{\sigma}\chi_{\beta}^{(+)}$ in the terms proportional to $(\tilde{b}_{i}-\delta_{nq}\tilde{b}_{i}^{\prime}) (i=0,3)$ in Eq.~(\ref{eq:inelastic f QRPA}). These terms are associated with the spin terms of the Skyrme force \cite{Vesely:2009eb}. The spin term could be expanded by three different terms $\lambda^{\pm,z}_{LM}(r)$ in the spherical coordinate,
\begin{widetext}
\begin{equation}
\begin{split}
\label{eq:spin expansion}
    \chi_{\alpha}^{(-)*}\vec{\sigma}\chi_{\beta}^{(+)}&=\frac{\vec{e}_{x}-i\vec{e}_{y}}{2}(\chi_{\alpha}^{(-)*}\sigma_{+}\chi_{\beta}^{(+)})+\frac{\vec{e}_{x}+i\vec{e}_{y}}{2}(\chi_{\alpha}^{(-)*}\sigma_{-}\chi_{\beta}^{(+)})+\vec{e}_{z}(\chi_{\alpha}^{(-)*}\sigma_{z}\chi_{\beta}^{(+)})\\ 
    &\equiv\sum_{LM}Y_{LM}(\theta,\phi)\left\{
    \frac{\vec{e}_{x}-i\vec{e}_{y}}{2}\lambda^{+}_{LM}(r)+\frac{\vec{e}_{x}+i\vec{e}_{y}}{2}\lambda^{-}_{LM}(r)+\vec{e}_{z}\lambda^{z}_{LM}(r)\right\},\\
\end{split}
\end{equation}
\end{widetext}
\begin{equation}
\label{eq:lambdams}
    \lambda^{m_s}_{LM}(r)=\int\mathrm{d}\Omega Y_{LM}^{*}(\theta,\phi)\chi_{\alpha}^{(-)*}\sigma_{m_s}\chi_{\beta}^{(+)}\ \ (m_s=\pm,z).
\end{equation}
The M$L$($L=1,2,\ \dots$) transition is induced by the vector of $\nabla(r^{L}Y_{LM}(\theta,\phi))$ that satisfies the orthogonality,
\begin{equation}
\int\mathrm{d}\Omega\frac{\nabla(r^{L^{\prime}}Y_{L^{\prime}M^{\prime}})^{*}\cdot\nabla(r^{L}Y_{LM})}{r^{2(L-1)}L(2L+1)}=\delta_{L^{\prime}L}\delta_{M^{\prime}M}.
\end{equation}
Therefore, the partial contribution from the $J^{\Pi}=L^{(-)^{L-1}}\ (L=1,2,\ \ldots)$ transition in Eq.~(\ref{eq:spin expansion}) is calculated from the inner product below,
\begin{equation}
\begin{split}
\label{eq:spin partial contribution Ml}
    \kappa_{LM}(r)&\equiv\int\mathrm{d}\Omega\frac{\nabla(r^{L}Y_{LM})^{*}\cdot\chi_{\alpha}^{(-)*}\vec{\sigma}\chi_{\beta}^{(+)}}{r^{2(L-1)}L(2L+1)}\\
    &=\frac{1}{2Lr^{L-1}}\sqrt{\frac{(L-M-1)(L-M)}{(2L+1)(2L-1)}}\lambda^{+}_{L-1,M+1}(r)\\
    &+\frac{1}{Lr^{L-1}}\sqrt{\frac{(L-M)(L+M)}{(2L+1)(2L-1)}}\lambda^{z}_{L-1,M}(r)\\
    &-\frac{1}{2Lr^{L-1}}\sqrt{\frac{(L+M)(L+M-1)}{(2L+1)(2L-1)}}\lambda^{-}_{L-1,M-1}(r).
\end{split}
\end{equation}
The orthogonality of the vector spherical harmonics such as $\{\vec{\Psi}_{lm},\vec{Y}_{lm},\vec{\Phi}_{lm}\}$ \cite{Barrera1985EJPh} is used for the above calculations. From Eqs.~(\ref{eq:spin expansion}) and (\ref{eq:spin partial contribution Ml}), the partial contribution from the unnatural parity transition in Eq.~(\ref{eq:inelastic f QRPA}) is calculated. For example, the term proportional to $(\tilde{b}_{0}-\delta_{nq}\tilde{b}_{0}^{\prime})$ in Eq.~(\ref{eq:inelastic f QRPA}) includes the contribution from the $J^{\Pi}=L^{(-)^{L-1}}\ (L=1,2,\ \ldots)$ transition, which is given by 
\begin{equation}
\label{eq:ML contribution e.g.}
    (\tilde{b}_{0}-\delta_{qq^{\prime}}\tilde{b}_{0}^{\prime})\int\mathrm{d}^{3}r\ \phi_{\mu}^{q*}\sum_{M}\kappa_{LM}(r)\vec{\sigma}\cdot\nabla(r^{L}Y_{LM})\phi_{\nu}^{q}.
\end{equation}
We remark that the product $\chi_{\alpha}^{(-)*}\vec{\sigma}\chi_{\beta}^{(+)}$ can induce natural parity transitions too as suggested in a microscopic calculation \cite{Johnson:1966zz}. We ignore such contributions of the natural parity transitions due to dominant contributions from the two-body and three-body interactions relevant to $b_i,b_i^{\prime}(i=0,3)$ in Eq.~(\ref{eq:inelastic f QRPA}). The partial contribution from the $J^{\Pi}$ transition is originated from the $\lambda$ coefficients in Eqs.~(\ref{eq:lambda0}) and (\ref{eq:lambdams}) composed of the product of distorted waves. These coefficients are easily calculated when the distorted waves are described in the spherical coordinate.

\subsection{Product of distorted waves}
\label{sec:distorted wave}
In the spherical coordinate $(r,\theta,\phi)$, the distorted wave used for the entrance channel with the momentum $\vec{k}$ and the spin $z$-component $\Sigma$ in $r\to\infty$ is given by \cite{SATCHLER19641},
\begin{eqnarray}
\chi^{(+)}_{\Sigma}&=&\sum_{m=\pm1/2}\chi^{(+)}_{m\Sigma}(\vec{r},\vec{k})\chi_{m}(\sigma),\label{eq:chi+}\\
\chi^{(+)}_{m\Sigma}(\vec{r},\vec{k})&=&\frac{4\pi}{kr}\sum_{JLM}C^{L,1/2,J}_{M,\Sigma,M+\Sigma}C^{L,1/2,J}_{M+\Sigma_-m,m,M+\Sigma}\nonumber\\
&\times&i^{L}\chi_{LJ}^{(+)}(k,r)Y_{LM}^{*}(\hat{k})Y_{LM+\Sigma-m}(\hat{r}),
\end{eqnarray}
where $\chi_{m}(\sigma)\ (m=\pm1/2)$ is the spin eigenstate of the incoming neutron. 
$C^{L,1/2,J}_{M,m,M+m}$represents
the Clebsch–Gordan coefficient for the angular momentum $J$ and its $z$-component $M+m$. $\hat{k}$ and $\hat{r}$ are unit vectors in the directions of $\vec{k}$ and $\vec{r}$. $\chi_{LJ}^{(+)}(k,r)$ is the radial part of the distorted wave obtained by solving the radial Schr$\ddot{\mathrm{o}}$dinger equation in the optical model. From the time-reversal symmetry, the spin component of the distorted wave used for the exit channel is given by
\cite{SATCHLER19641},
\begin{eqnarray}
\label{eq:chi- components}
\chi^{(-)}_{m\Sigma}(\vec{r},\vec{k})^{*}=(-1)^{\Sigma-m}\chi^{(+)}_{-m-\Sigma}(\vec{r},-\vec{k}).
\end{eqnarray}
Without loss of generality, we can assume that the direction of incoming neutron is along the $z$-axis ($\vec{k}_{\beta}\parallel \hat{z}$) and a neutron is emitted on the $x$--$z$ plane ($\vec{k}_{\alpha}\times\vec{k}_{\beta}\parallel \hat{y}$ or $\phi_{k_\alpha}=0$) for the unpolarized reaction. This assumption reduces $\lambda^{0}_{LM}$ in Eq.~(\ref{eq:lambda0}) to a function of the scattering angle $\theta_{k_\alpha}\in [0,\pi]$ rad. From the expressions of the distorted waves in Eqs.~(\ref{eq:chi+})-(\ref{eq:chi- components}), the $\lambda^{0}_{LM}$ in Eq.~(\ref{eq:lambda0}) is given by
\begin{eqnarray}
\label{eq:lambda0v2}
    \lambda^{0}_{LM}(r)&=&\int\mathrm{d}\Omega\sum_{m}Y_{LM}^{*}(\theta,\phi)\chi^{(-)}_{m\Sigma_{\alpha}}(\vec{r},\vec{k}_{\alpha})^{*}\chi^{(+)}_{m\Sigma_{\beta}}(\vec{r},\vec{k}_{\beta})\nonumber\\
    &=&\sum_{J_{\alpha L_{\alpha}}}\sum_{J_{\beta}L_{\beta}}\sum_{m}(-1)^{\Sigma_{\alpha}-m}\sqrt{\frac{(2L_{\alpha}+1)(2L_{\beta}+1)}{4\pi(2L+1)}}
    \nonumber\\
    &\times&C^{L_{\alpha},\frac{1}{2},J_{\alpha}}_{M_{\alpha},-\Sigma_{\alpha},M_{\alpha}-\Sigma_{\alpha}}C^{L_{\alpha},\frac{1}{2},J_{\alpha}}_{M_{\alpha}-\Sigma_{\alpha}+m,-m,M_{\alpha}-\Sigma_{\alpha}}\nonumber\\
    &\times&C^{L_{\beta},\frac{1}{2},J_{\beta}}_{0,\Sigma_{\beta},\Sigma_{\beta}}C^{L_{\beta},\frac{1}{2},J_{\beta}}_{\Sigma_{\beta}-m,m,\Sigma_{\beta}}\nonumber\\
    &\times&C^{L_{\alpha},L_{\beta},L}_{M_{\alpha}-\Sigma_{\alpha}+m,\Sigma_{\beta}-m,M}C^{L_{\alpha},L_{\beta},L}_{0,0,0}\nonumber\\
    &\times&\frac{(4\pi)^2}{k_{\alpha}k_{\beta}r^{2}} \chi_{L_{\alpha}J_{\alpha}}^{(+)}(k_{\alpha},r)
    \chi_{L_{\beta}J_{\beta}}^{(+)}(k_{\beta},r)\nonumber\\
    &\times&i^{L_{\alpha}+L_{\beta}}Y^{*}_{L_{\alpha}M_{\alpha}}(-\hat{k}_{\alpha})|_{\phi_{k_\alpha}=0}\sqrt{\frac{2L_{\beta}+1}{4\pi}},
\end{eqnarray}
where $M_{\alpha}=M+\Sigma_{\alpha}-\Sigma_{\beta}$ and the integration formula of spherical harmonics,
\begin{eqnarray}
    &{\displaystyle \int }&\mathrm{d}\Omega Y_{LM}^{*}(\theta,\phi)Y_{L_\alpha M_{\alpha}}(\theta,\phi)Y_{L_\beta M_\beta}(\theta,\phi)\nonumber\\
    &=&\sqrt{\frac{(2L_{\alpha}+1)(2L_{\beta}+1)}{4\pi(2L+1)}}C^{L_{\alpha},L_{\beta},L}_{M_{\alpha},M_{\beta},M}C^{L_{\alpha},L_{\beta},L}_{0,0,0},
\end{eqnarray}
is used for the derivation. Other $\lambda$ coefficients in Eq.~(\ref{eq:lambdams}) used for Eq.~(\ref{eq:spin partial contribution Ml}) are calculated in the same way as Eq.~(\ref{eq:lambda0v2}), and the detail equations are shown in the Appendix~\ref{sec:lambda unnatural parity}.

\subsection{Spin distribution}
\label{sec:spin parameters}
Microscopically calculated inelastic scattering process provides the distribution of populated spin states in the residual nucleus \cite{Kerveno2021}. The spin distribution of the residual nucleus is characterized by the excitation energy $E_x$, the angular momentum $J$, and the outgoing neutron energy $E_\mathrm{out}(=\hbar^2k_\alpha^2/2\mu_\alpha)$. The spin distribution after the pre-equilibrium reaction of $(n,n^{\prime})$ for a given neutron incident energy is calculated from the angular integration of Eq.~(\ref{eq:transition strength QRPA+DWBA}),
\begin{equation}
\label{eq:spin dis}
R(E_{\mathrm{out}},E_x,J)=\frac{\sum_{\Sigma_\alpha\Sigma_\beta M\Pi}\int\mathrm{d}\Omega_\alpha\frac{\mathrm{d}B(\omega;F^{M}_{\mathrm{inl}}(J^{\Pi}))}{\mathrm{d}E_x}}{\sum_{J}\sum_{\Sigma_\alpha\Sigma_\beta M\Pi}\int\mathrm{d}\Omega_\alpha\frac{\mathrm{d}B(\omega;F^{M}_{\mathrm{inl}}(J^{\Pi}))}{\mathrm{d}E_x}},
\end{equation}
where 
$E_{\mathrm{out}}=E_{\mathrm{in}}-E_x$ for a incident neutron energy $E_{\mathrm{in}}(=\hbar^2k_\beta^2/2\mu_\beta)$ in the center of mass system. 


\section{Results and Discussions}
\label{sec:Results and Discussions}

\subsection{Excitations to low-lying discrete levels}
\label{sec:discrete levels}

\begin{table}
\caption{Comparison of the QRPA results with the evaluated values \cite{Martin20071583} for the strengths and the excitation energies of low-lying discrete levels in $^{208}$Pb. The units of the strengths and excitation energies are $e^{2}b^{L}$ and MeV, respectively, for the E$L$ transitions ($L=2,3,5$).}
    \centering
    \begin{tabular}{c|c c c c}\hline\hline
Level& $B_{\mathrm{exp}}(\mathrm{E}L)$ & $B_{\mathrm{QRPA}}(\mathrm{E}L)$& $E_\mathrm{exp}$ &  $E_\mathrm{QRPA}$ \\
    \hline
        $3^{-}_{1}$& 0.611& 0.656 &2.615 &3.625 \\
        $5^{-}_{1}$ & 0.0447 &0.0415 &3.198&4.625 \\
        $2^{+}_{1}$ & 0.318 &0.298&4.086&4.938 \\
        \hline
    \end{tabular}
    \label{tab:strength data}
\end{table}

\begin{figure}[t]
\includegraphics[width=1\linewidth]{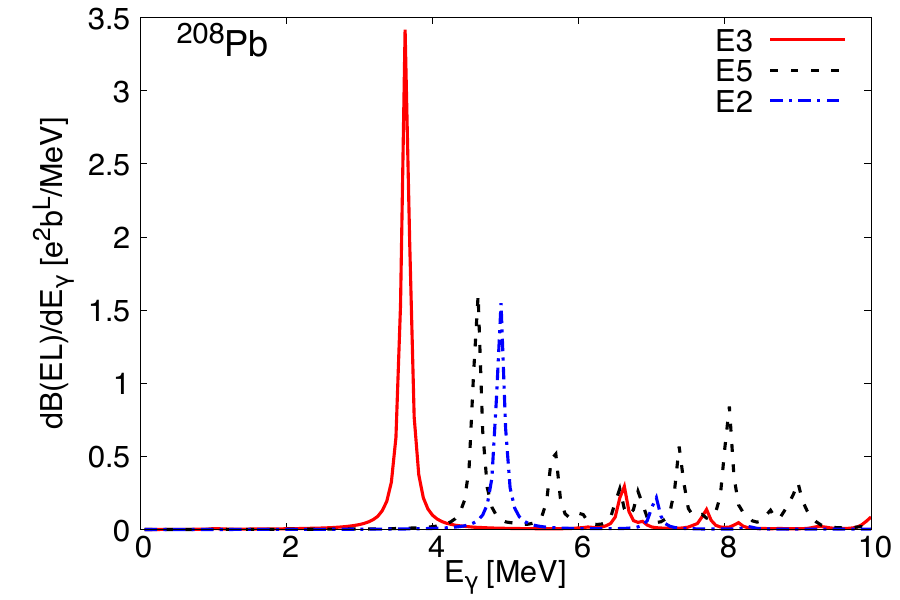}
\caption{
Strength functions of E2 (dash-dotted line), E3 (solid line), and E5 (dashed line) transitions for photoabsorption reactions on $^{208}$Pb, which is calculated from the QRPA equation in Eq.~(\ref{eq:QRPAeq}) with the external fields of E$L$ operators.
}
\label{fig:strength}
\end{figure}

\begin{figure}[h]
\includegraphics[width=0.8\linewidth]{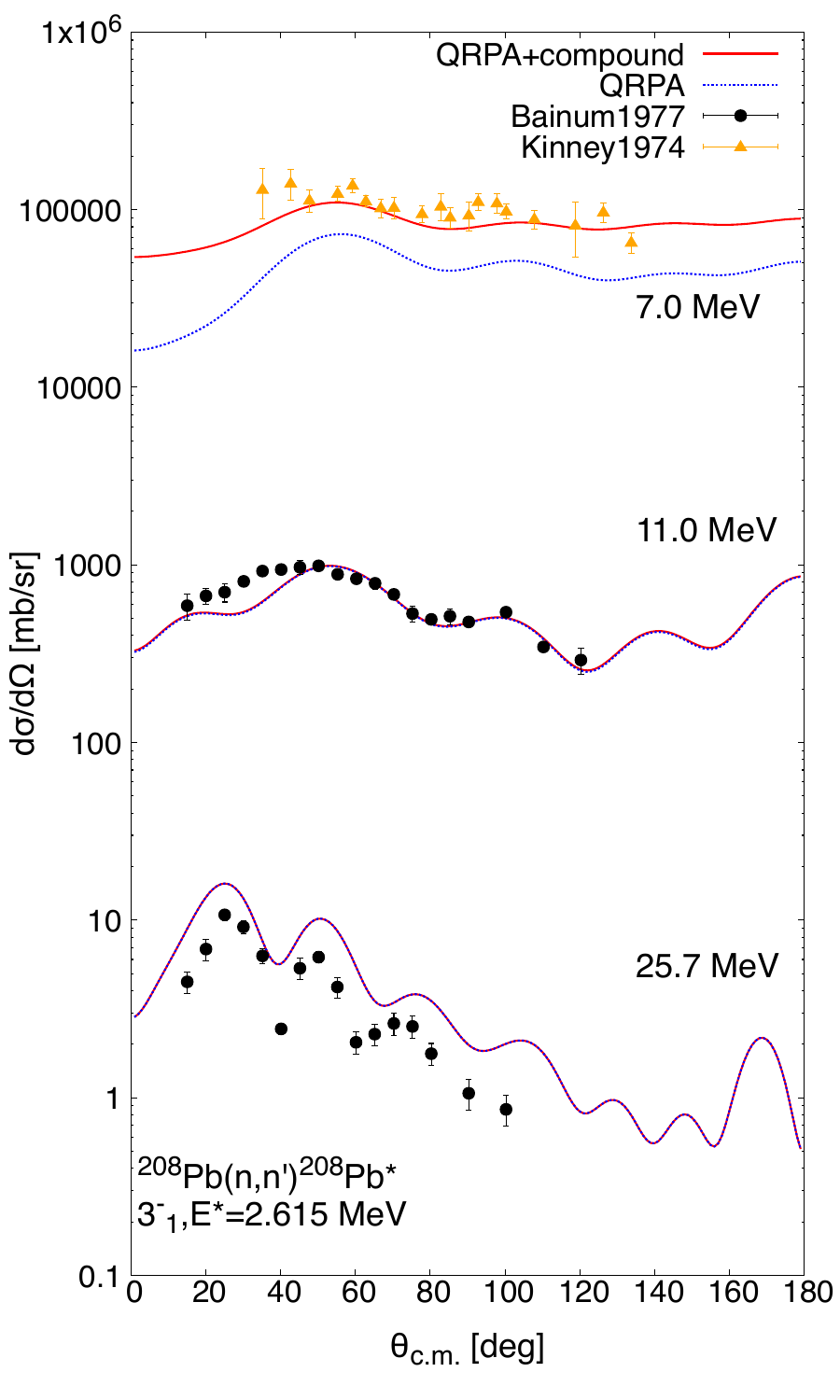}
\caption{
The differential cross sections for the neutron-induced inelastic scattering on a $^{208}$Pb target, exciting into 
the $3^{-}_1(2.615\mathrm{MeV})$ state. The dotted lines represent the results of Eq.~(\ref{eq:differential cs discrete}) for three neutron incident energies: 7.0, 11.0, and 25.7 MeV. The solid lines show the sum of contributions from Eq.~(\ref{eq:differential cs discrete}) and the compound process. The filled triangles and the filled circles represent experimental data from Refs.~\cite{Kinney_Perey_1974,Bainum1977}. Both the data and calculated results for 7.0 and 11.0 MeV are scaled by $10^{4}$ and $10^{2}$ times, respectively. 
}
\label{fig:3-}
\end{figure}

\begin{figure}[h]
\includegraphics[width=0.8\linewidth]{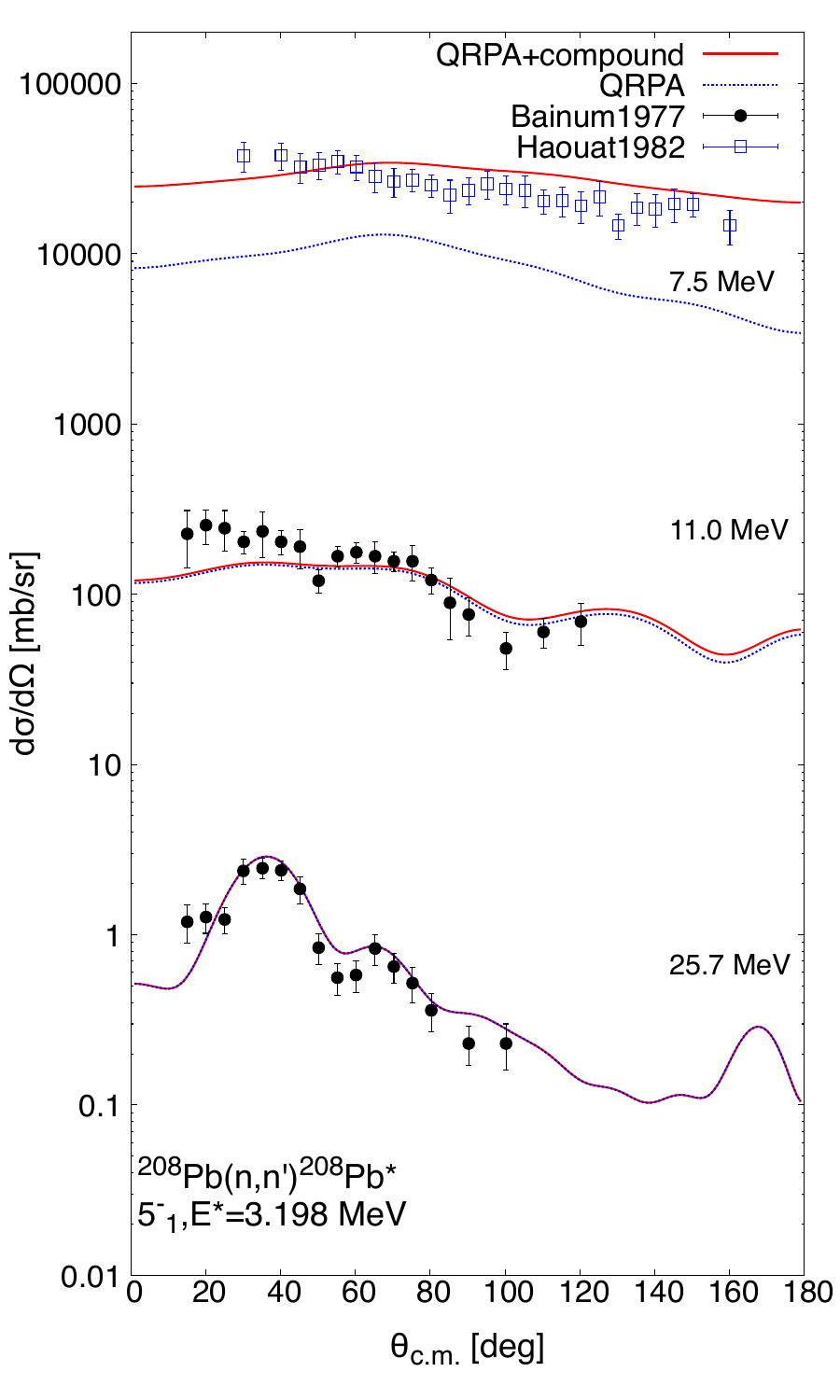}
\caption{
The differential cross sections for excitation to the $5^{-}_1(3.198\ \mathrm{MeV})$ state at incident neutron energies of 7.5, 11.0, and 25.7 MeV. The dotted and solid lines are calculated in the same way as those in Fig.~\ref{fig:3-}. The open squares and the filled circles represent experimental data from Refs.~\cite{Haouat1982r2200,Haouat1982r2284,Dupuis2019,Bainum1977}. The calculated results and experimental data for 7.5 and 11.0 MeV are scaled by factors of $10^{4}$ and $10^{2}$.
}
\label{fig:5-}
\end{figure}

\begin{figure}
\includegraphics[width=0.8\linewidth]{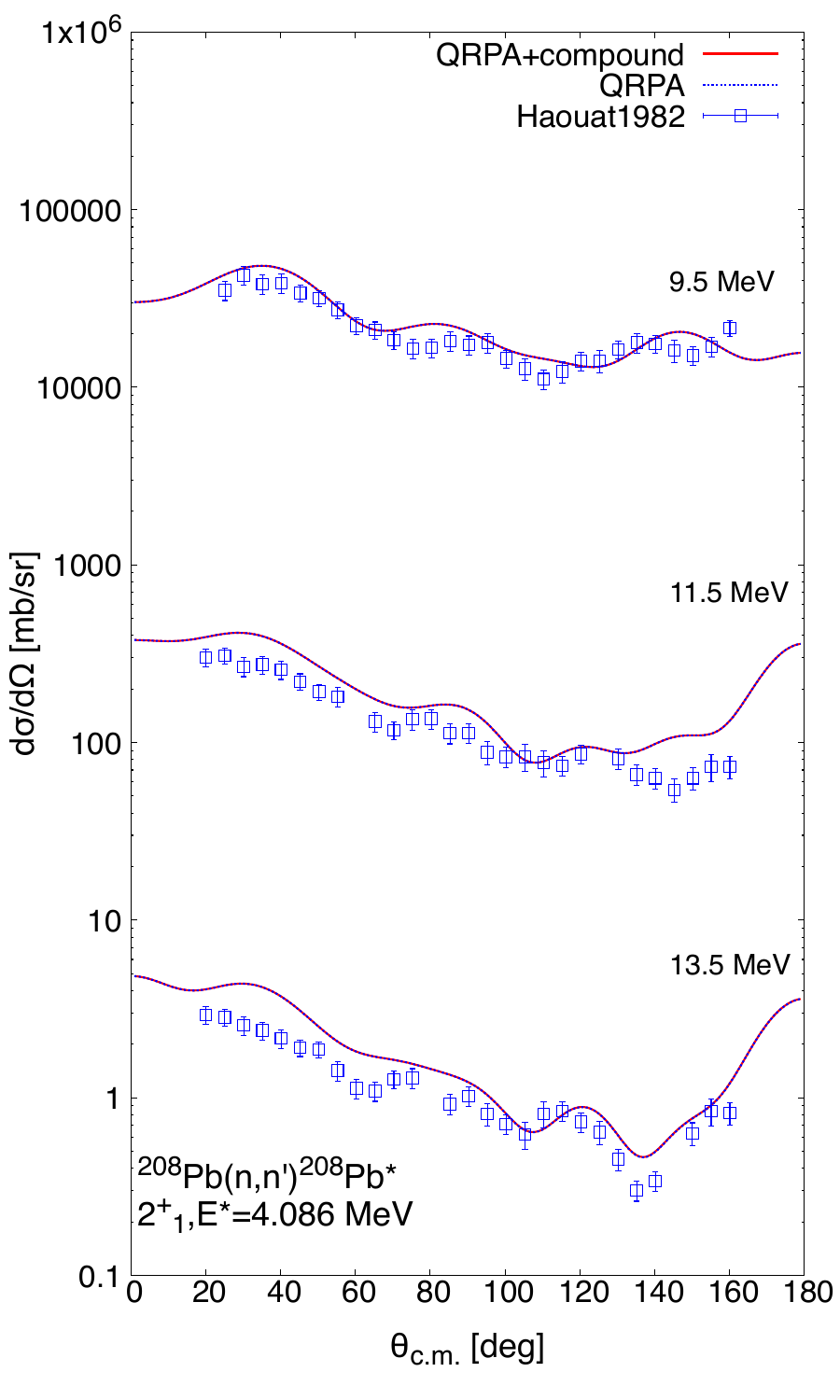}
\caption{
The calculated results and experimental data \cite{Haouat1982r2200,Haouat1982r2284,Dupuis2019} for excitation to 
the $2^{+}_1(4.086\ \mathrm{MeV})$ state at incident energies of 9.5 MeV, 11.5 MeV, and 13.5 MeV, as in Figs.~\ref{fig:3-} and \ref{fig:5-}.
}
\label{fig:2+}
\end{figure}

We solve the QRPA equation with the external field of the neutron-induced direct inelastic scattering on $^{208}$Pb target following the method in Sec.~\ref{sec:theory}. The single-particle states are obtained from the HF+BCS calculation  in the cylindrical coordinate by using the numerical setup in Ref.~\cite{Bonneau:2007dc}. Products of the single-particle states, such as $\phi_{\mu}^{q*}\phi_{\nu}$, $\nabla\phi_{\mu}^{q*}\cdot\nabla\phi_{\nu}^{q}$, $\nabla^{2}(\phi_{\mu}^{q*}\phi_{\nu}^{q})$, $\nabla\phi_{\mu}^{q*}\cdot\left(
    -i
    \right)\left(
    \nabla\times\vec{\sigma}
    \right)\phi_{\nu}^{q}$, and $\phi_{\mu}^{q*}\vec{\sigma}\phi_{\nu}^{q}$ in Eq.~(\ref{eq:inelastic f QRPA})
are calculated as in Ref.~\cite{Sasaki:2022ipn}. We employ the Skyrme parameters of SLy4 \cite{Chabanat:1997un} for both the QRPA matrices in Eq.~(\ref{eq:QRPAeq}) and the external field in Eq.~(\ref{eq:inelastic f QRPA}). The size of the configuration space of quasiparticle pairs is determined following Ref.~\cite{SasakiPRC2023}. To obtain the distorted waves of the incoming and outgoing neutrons, we solve the optical model using the coupled-channels Hauser-Feshbach code CoH$_3$ \cite{Kawano2021}. The calculation employs the neutron optical potential parameters from Ref.~\cite{Kunieda2007}, and includes the non-local Perey effect ($\beta=0.85$) \cite{Perey1964}. \\
\indent For the benchmark calculation, we demonstrate neutron-induced direct inelastic scattering to the low-lying discrete levels of $^{208}$Pb at various incident neutron energies. To calculate the inelastic scattering to these discrete levels, first we calculate the excitation energies by the QRPA for the photoabsorption reaction. Later these calculated excitation energies will be adjusted to the experimental values. Figure~\ref{fig:strength} shows the transition strengths of the E2, E3, and E5 transitions for $^{208}$Pb obtained from the QRPA equation in Eq.~(\ref{eq:QRPAeq}) with the external fields of $e(\frac{1}{2}-\tau^{q}_z)r^{L}Y_{LM}(L=2,3,5,\ |M|\leq L)$ as given in Eq.~(\ref{eq:external field f}). The photon energy grid is set to 62.5 keV, and the strength is calculated from 62.5 keV to 20 MeV using a fixed Lorentzian width $\gamma=2\mathrm{Im}(\omega)=125$ keV as in Ref.~\cite{SasakiPRC2023}. The prominent resonances at low energies correspond to 
excitations to low-lying states in $^{208}$Pb, such as  $3^{-}_1(2.615\ \mathrm{MeV})$, $5^{-}_1(3.198\ \mathrm{MeV})$, and 
$2^{+}_1(4.086\ \mathrm{MeV})$. Our QRPA calculations overestimate the excitation energy by at most 1.5 MeV. Such overestimation is also seen in a previous study \cite{Dupuis:2011zza}. The transition strengths of the low energy resonances are calculated by integrating over the energy region around the peak, as described in Eq.~(\ref{eq:differential cs discrete}), with $\Delta=1$ MeV for the E3 and E2 transitions, and $\Delta=0.5$ MeV for the E5 transition. The obtained values shown in Table~\ref{tab:strength data} are consistent with the adopted values \cite{Martin20071583} within $10\%$. \\
\indent For the excitation to a discrete level, the differential cross section is calculated using $E_x=E_{\mathrm{{exp}}}$ in Eq.~(\ref{eq:differential cs discrete}) where $E_{\mathrm{{exp}}}$ is the evaluated value of the excitation energy in Table~\ref{tab:strength data}. To correct the overestimation of the excitation energy, we introduce the energy shift $\Delta_{N}$ \cite{Dupuis:2011zza}. This correction is implemented by modifying Eq.~(\ref{eq:QRPAeq}) with the replacement $\omega\to\omega+\Delta_{N}$. The value of the energy shift is defined as $\Delta_{N}=E_{\mathrm{QRPA}}-E_{\mathrm{exp}}$, where $E_{\mathrm{QRPA}}$ is the QRPA-calculated excitation energy as in Table~\ref{tab:strength data}.

Figure~\ref{fig:3-} shows the differential cross sections for excitation to the $3^{-}_1(2.615\ \mathrm{MeV})$ state at incident neutron energies of 7.0 MeV, 11.0 MeV, and 25.7 MeV. The results for 7.0 MeV and 11.0 MeV are scaled by $10^{4}$ and $10^{2}$ times. The dotted lines in Fig.~\ref{fig:3-} represent the differential cross sections calculated from Eq.~(\ref{eq:differential cs discrete}), corresponding to the direct reaction contribution. The solid lines show the sum of the contributions from direct reactions and compound nucleus decays. The compound process is calculated using CoH$_3$ \cite{Kawano2021}, and its contribution dominates at low incident neutron energies. At the 7.0-MeV neutron incident, the sum of the direct reaction process of Eq.~(\ref{eq:differential cs discrete}) and the compound nuclear reaction agrees with the experimental data. While the compound reaction contribution is negligible for the 11.0 and 25.7~MeV cases. The calculated result for 11.0 MeV reproduces both the shape and strength of the experimental data. The strength is entirely  determined by the values of the Skyrme $b$ parameters in Eq.~(\ref{eq:inelastic f QRPA}), without any fitting of deformation parameters for vibrations, as is typically done in the collective model within the DWBA framework. The result for 25.7 MeV reproduces the shape of the distribution, which is similar to the result in Ref.~\cite{Dupuis2019}. The overestimation of the strength could be improved by including various terms in Eq.~(\ref{eq:inelastic f QRPA}) that are neglected in this study and by using a smaller parameter $\beta$ for the non-local Perey effect.

\begin{figure}
\includegraphics[width=1\linewidth]{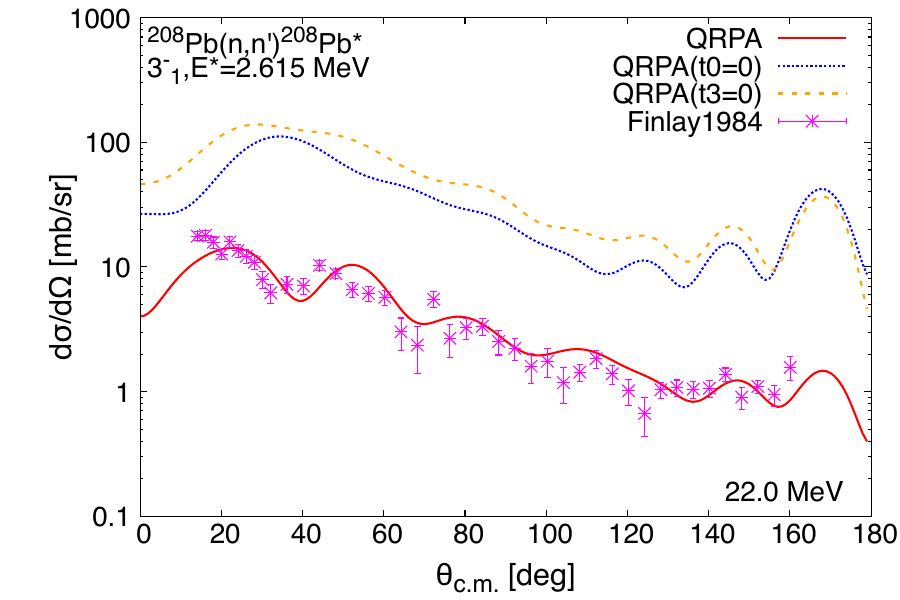}
\caption{
The results of Eq.~(\ref{eq:differential cs discrete}) for the $^{208}\mathrm{Pb}(n,n^{\prime})^{208}\mathrm{Pb}^{*}(3^{-}_1)$ reaction at a 22.0-MeV incident neutron. The dotted (dashed) line shows the result obtained by assuming $t_{0}=0  (t_{3}=0)$ in the Skyrme force of Eq.~(\ref{eq:skyrme}). The solid line is calculated with the finite values of both $t_{0}$ and $t_{3}$ in the SLy4 parameter set. The symbols represent the experimental data \cite{Finlay:1984prc}.
}
\label{fig:3-_Skyrme parameter}
\end{figure}

Figures~\ref{fig:5-} and \ref{fig:2+} show the differential cross sections for excitation to  the $5^{-}_1(3.198\ \mathrm{MeV})$ and 
$2^{+}_1(4.086\ \mathrm{MeV})$ states. As seen in Fig.~\ref{fig:5-}, the compound inelastic scattering process is larger than the direct process at 7.5 MeV. The calculated results of Eq.~(\ref{eq:differential cs discrete}) reproduce well both the shapes and strengths of the differential cross sections at 25.7 MeV in Fig.~\ref{fig:5-} and 9.5 MeV in Fig.~\ref{fig:2+}.

Figure~\ref{fig:3-_Skyrme parameter} demonstrates the sensitivity to two-body and three-body Skyrme forces, which are proportional to $t_0$ and $t_3$ in Eq.~(\ref{eq:skyrme}). At an incident neutron energy of 22.0 MeV, the contribution from the compound process is negligible. The dotted line shows the result of Eq.~(\ref{eq:differential cs discrete}) when the two-body force is neglected ($t_0=0$), which corresponds to $b_{0}=b_{0}^{\prime}=0$
in Eq.~(\ref{eq:inelastic f QRPA}). 
Similarly, the dashed line is the case when we omit the three-body force ($t_3=b_{3}=b_{3}^{\prime}=0$). Both the dotted and dashed lines overestimate the strength by an order of magnitude. In contrast, the result with finite values of $t_0$ and $t_3$ (solid line) is smaller than both of the cases, and the calculated cross section agrees well with the experimental data \cite{Finlay:1984prc}. This strong cancellation between the two-body and three-body forces is also reported by Davies and Satchler for $^{40}$Ca with Skyrme forces \cite{Davies:1974ell}. The factor of $(\alpha+2)(\alpha+1)/2$ in Eq.~(\ref{eq:inelastic f QRPA}) is essential for achieving this cancellation. Additionally, a smaller non-local Perey factor may help better agree with experimental data.

\begin{figure*}[t]
\subfigure{%
    \includegraphics[clip, width=0.9\columnwidth]{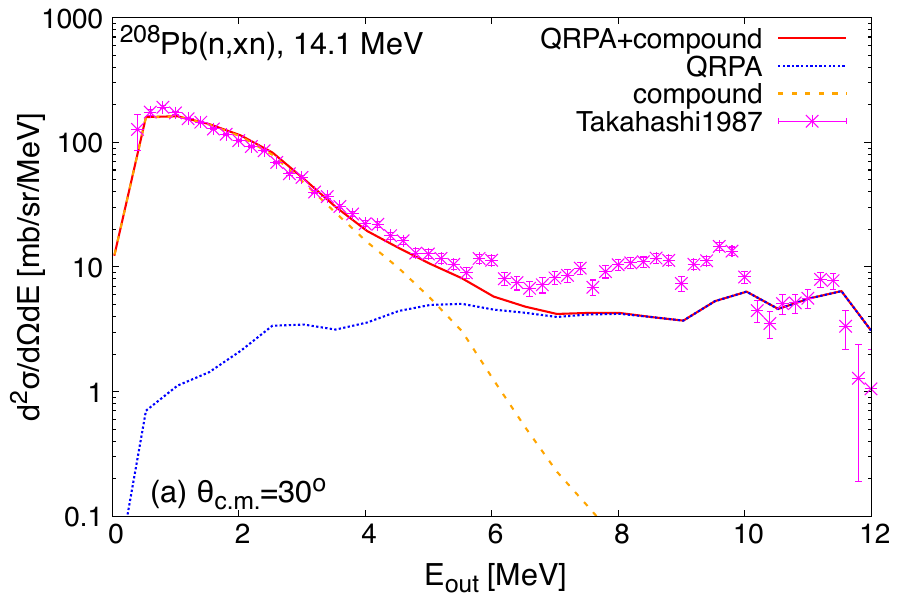}}%
\subfigure{%
    \includegraphics[clip, width=0.9\columnwidth]{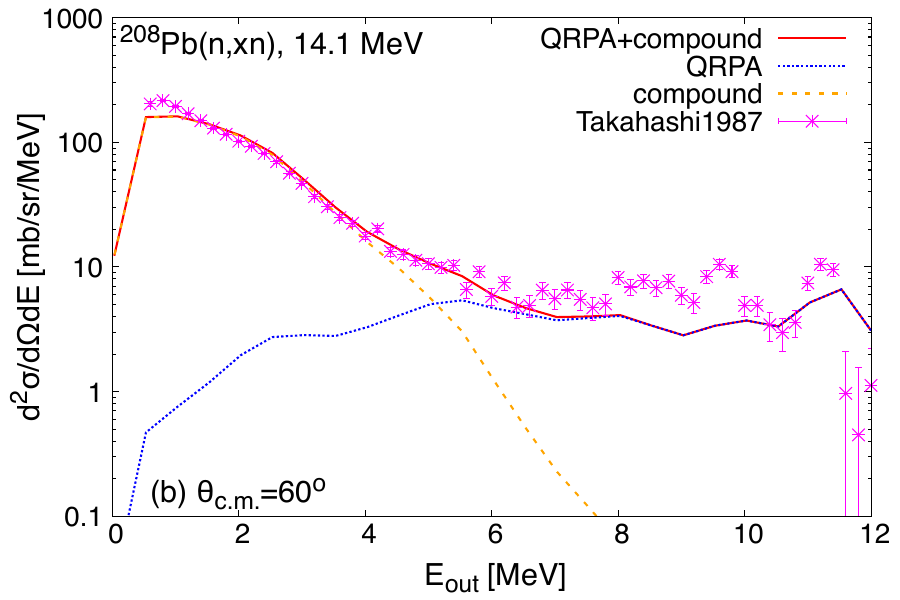}}%
\end{figure*}
\begin{figure*}
\subfigure{%
    \includegraphics[clip, width=0.9\columnwidth]{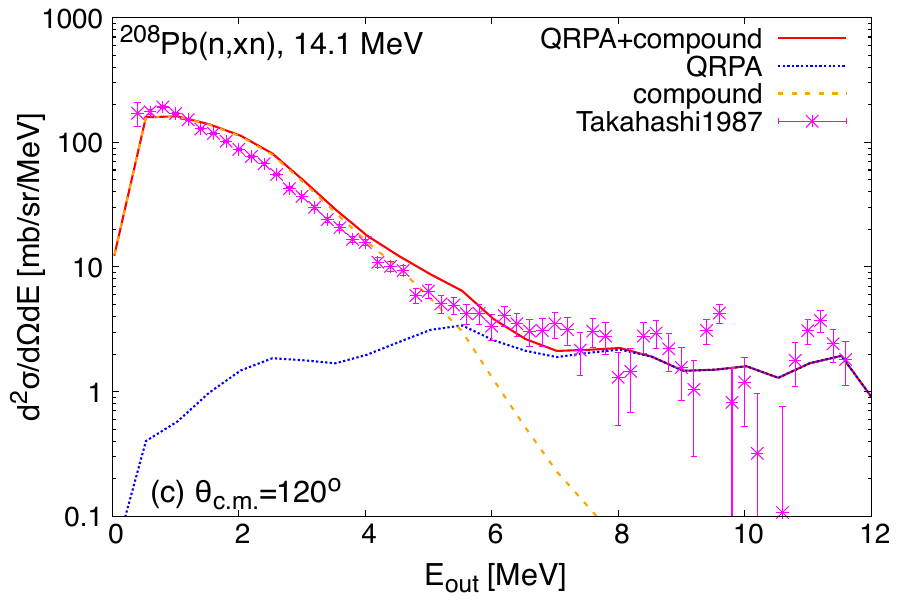}}%
\subfigure{%
    \includegraphics[clip, width=0.9\columnwidth]{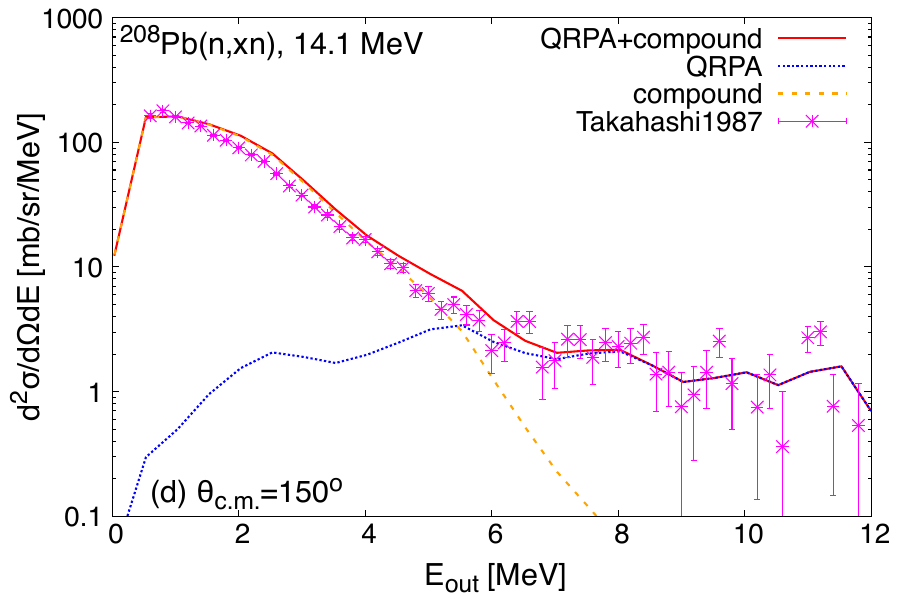}}%
\caption{The double differential cross sections for neutron--induced reaction on $^{208}$Pb with a 14.1-MeV incident neutron at the neutron scattering angles: (a) $\theta_{\mathrm{c.m.}}=30^{\mathrm{o}}$, (b) $\theta_{\mathrm{c.m.}}=60^{\mathrm{o}}$, (c) $\theta_{\mathrm{c.m.}}=120^{\mathrm{o}}$ and (d) $\theta_{\mathrm{c.m.}}=150^{\mathrm{o}}$. The dotted lines show the results of Eq.~(\ref{eq:double differential cs}). The dashed lines show the compound nucleus decay contributions. The solid lines represent the sum of both the dotted and dashed lines. The symbols represent the experimental data \cite{Takahashi01031988}.}
\label{fig:ddw}
\end{figure*}

\subsection{Excitations to continuum states}
\label{sec:continuum}

We calculate the double differential cross section given in Eq.~(\ref{eq:double differential cs}) considering both natural parity transitions ($0^{+},1^{-},2^{+},..,10^{+}$) and unnatural parity transitions ($1^{+},2^{-},3^{+},..,10^{-}$) up to $J=10$. The partial contributions from each $J^{\Pi}$ in Eq.~(\ref{eq:double differential cs}) are calculated following the decomposition of multipole transitions described in Sec.~\ref{sec:multipole}. The numerical setups for the single-particle state, the distorted wave, and the Skyrme force are the same as those in Sec.~\ref{sec:discrete levels}. The calculation is performed by increasing the excitation energy of the residual nucleus $E_{x}=\mathrm{Re}(\omega)$ from 0.5 MeV to the incident neutron energy at every 0.5 MeV with the Lorentzian width $\gamma=2\mathrm{Im}(\omega)=1.0$ MeV. 
The wave number of the outgoing neutron is determined from $k_{\alpha}=\sqrt{\mu_\alpha/\mu_\beta k_\beta^{2}-2\mu_\alpha E_{x}/\hbar^{2}}$. 
For the calculations of the ${2}^{+}, {3}^{-}, 4^{+},$ and ${5}^{-}$ transitions, we consider the energy shift of the transition strength, as done in Sec.~\ref{sec:discrete levels}, by correcting $\omega\to\omega+\Delta_{N}$ on the left-hand side of Eq.~(\ref{eq:QRPAeq}). A large spurious mode associated with the translational motion of the center of mass appears at low excitation energy in the $1^{-}$ transition. We remove this spurious mode following the prescription for a Hermitian operator in Ref.~\cite{SasakiPRC2023},  by changing the coefficient in Eq.~(\ref{eq:lambda0}): $\lambda^{0}_{1\pm1}\to\pm(\lambda^{0}_{1}+\lambda^{0}_{-1})/2$.

Figure~\ref{fig:ddw} shows the double differential cross sections of $^{208}\mathrm{Pb}(n,xn)$ for a 14.1-MeV incident neutron at two neutron scattering angles ($\theta_{\mathrm{c.m.}}=30^{\mathrm{o}},60^{\mathrm{o}}, 120^{\mathrm{o}}, \mathrm{and} \ 150^{\mathrm{o}}$). The dotted lines represent the results of Eq.~(\ref{eq:double differential cs}), which include contributions from direct- and pre-equilibrium processes. The dashed lines show the calculated results for the compound state decay contribution with CoH$_3$ \cite{Kawano2021}. The solid lines are the sum of both the dotted and dashed lines.\\
\indent The dotted lines coincide with the solid lines for $E_{\mathrm{out}}>8$ MeV, where the direct- and pre-equilibrium processes dominate. In Fig.~\ref{fig:ddw} panel (a) and (b), the dotted lines underestimate the experimental data by $20\%\mathrm{-}80\%$ in the region of $E_{\mathrm{out}}=8\mathrm{-}10 \ \mathrm{MeV}$. This underestimation could be improved by considering the neglected terms in Eq.~(\ref{eq:inelastic f QRPA}), using different Skyrme force parameters, or employing a larger configuration space in the QRPA matrices. On the other hand, in the case of the backward scatterings as shown in Fig.~\ref{fig:ddw} panel (c) and (d), the dotted lines are more consistent with the experimental data for $E_{\mathrm{out}}>8$ MeV.\\
\indent The contribution from the compound nucleus decay and that of the pre-equilibrium process are comparable in the region of $E_{\mathrm{out}}=4\mathrm{-}8 \ \mathrm{MeV}$. The sum of both contributions (solid line) reproduces the experimental data around 5 MeV in Fig.~\ref{fig:ddw} panel (a) and (b). In the low energy region, $E_{\mathrm{out}}< 4$ MeV, the dotted lines become much smaller than the dashed lines in Fig.~\ref{fig:ddw} because the level density of the residual nucleus increases exponentially as the outgoing neutron energy decreases, while the number of 1p-1h configuration is relatively insensitive to the excitation energy of the residual nucleus. There is also a contribution from the $(n,2n)$ reaction at low emission energies.



\begin{figure}[t]
\includegraphics[width=1\linewidth]{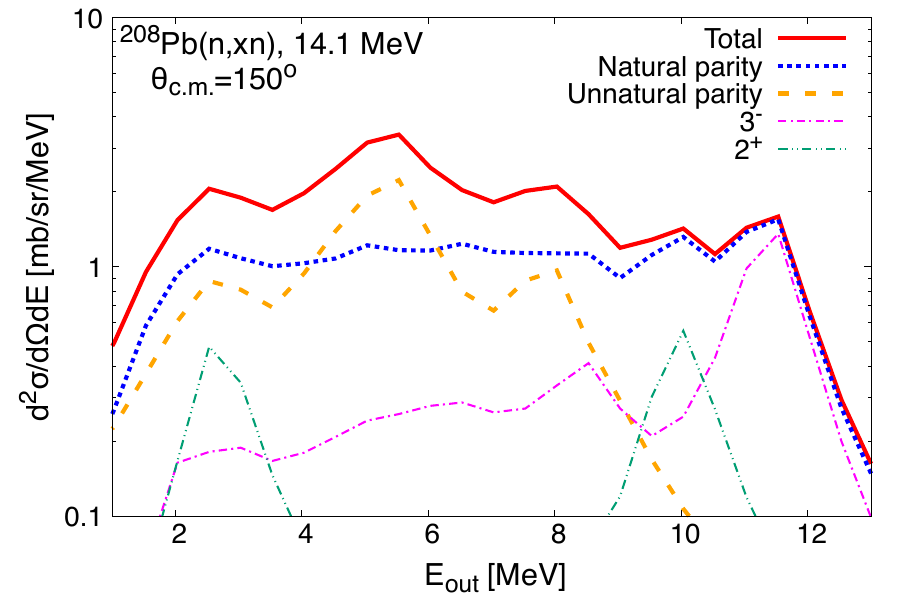}
\caption{
Partial contributions from different $J^{\pi}$ in the direct and pre-equilibrium processes of $^{208}\mathrm{Pb}(n,xn)$ with a 14.1-MeV incident neutron at $\theta_{\mathrm{c.m.}}=150^{\mathrm{o}}$. The solid line represents the result of Eq.~(\ref{eq:double differential cs}), including both natural and unnatural parity transitions, corresponding to the line of ``QRPA" in Fig.~\ref{fig:ddw}(d). The dotted line shows the result of Eq.~(\ref{eq:double differential cs}) considering only natural parity transitions ($J^{\pi}=0^{+},1^{-},..,10^{+}$). The dashed line shows the result considering only unnatural parity transitions ($J^{\pi}=1^{+},2^{-},..,10^{-}$). In addition, partial contributions from the $2^{+}$ and $3^{-}$ transitions to Eq.~(\ref{eq:double differential cs}) are also plotted in the figure.
}
\label{fig:ddw contr}
\end{figure}

Figure~\ref{fig:ddw contr} shows the 
results of Eq.~(\ref{eq:double differential cs}) for $150^{\mathrm{o}}$ at a 14.1-MeV incident neutron energy, along with the partial contributions from various $J^{\pi}$ states. The partial contributions at other neutron scattering angles are similar to those shown in Fig.~\ref{fig:ddw contr}. Contributions from excitations to the low-lying discrete levels, as in Table~\ref{tab:strength data}, are found in the region of $E_{\mathrm{out}}>9$ MeV. The partial contribution from the $3^{-}$ transition is the largest in the outgoing energy range $E_{\mathrm{out}}=11\mathrm{-}12$ MeV, which corresponds to the direct excitation of the $3^{-}_1(2.615\ \mathrm{MeV})$ state. The sharp peak around $E_{\mathrm{out}}=$10 MeV in the $2^{+}$ result represents excitation to the $2^{+}_1(4.086\ \mathrm{MeV})$ state. The contribution from natural parity transitions (dotted line) to the direct process in the region $E_{\mathrm{out}}=9\mathrm{-}12$ MeV is much larger than that from unnatural parity transitions (dashed line), which is also seen in Ref.~\cite{Dupuis:2011zza,dupuis2017microscopic}. The resonance around $E_{\mathrm{out}}=3$ MeV in the $2^{+}$ transition is induced by the isoscalar giant quadrupole resonance (ISGQR), which was observed experimentally \cite{harakeh2001giant} and reproduced numerically by the RPA calculations \cite{RING1973477,Roca-MazaPRC2013}. The unnatural parity transitions (dashed line) make a significant contribution to the pre-equilibrium process around $E_{\mathrm{out}}=5$ MeV. In particular, these transitions are necessary to reproduce the experimental data for $E_{\mathrm{out}}=4\mathrm{-}6$ MeV in Figs.~\ref{fig:ddw}(a) and \ref{fig:ddw}(b).

\subsection{Spin of the residual nucleus}
\label{sec:spin parameters}

\begin{figure}[t]
\includegraphics[width=1\linewidth]{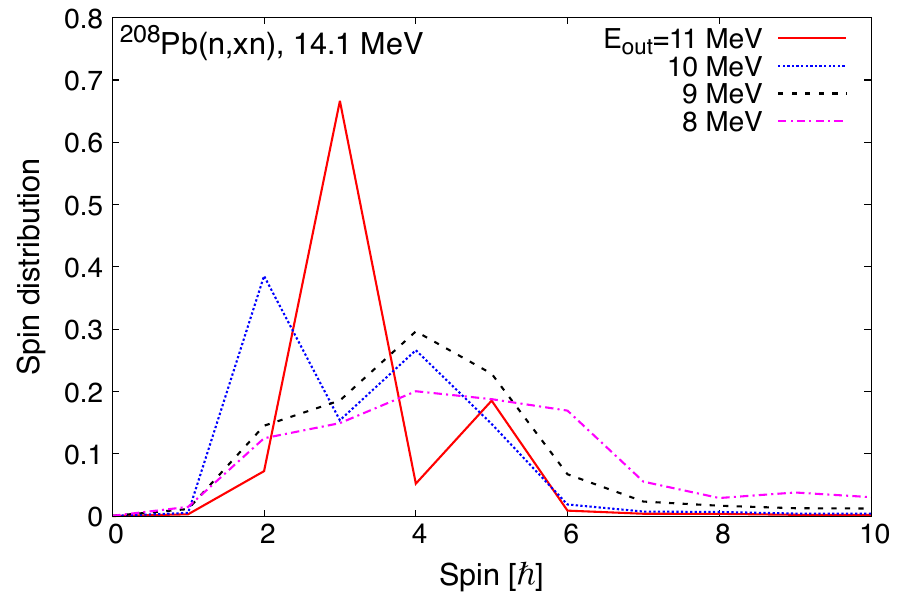}\\
\caption{
The spin distributions of the residual nucleus for different emitted neutron energies through the direct and one-step pre-equilibrium processes of $^{208}\mathrm{Pb}(n,xn)$ with a 14.1-MeV incident neutron, calculated from Eq.~(\ref{eq:spin dis}). 
}
\label{fig:spin distribution}
\end{figure}

\begin{figure}[t]
\includegraphics[width=1\linewidth]{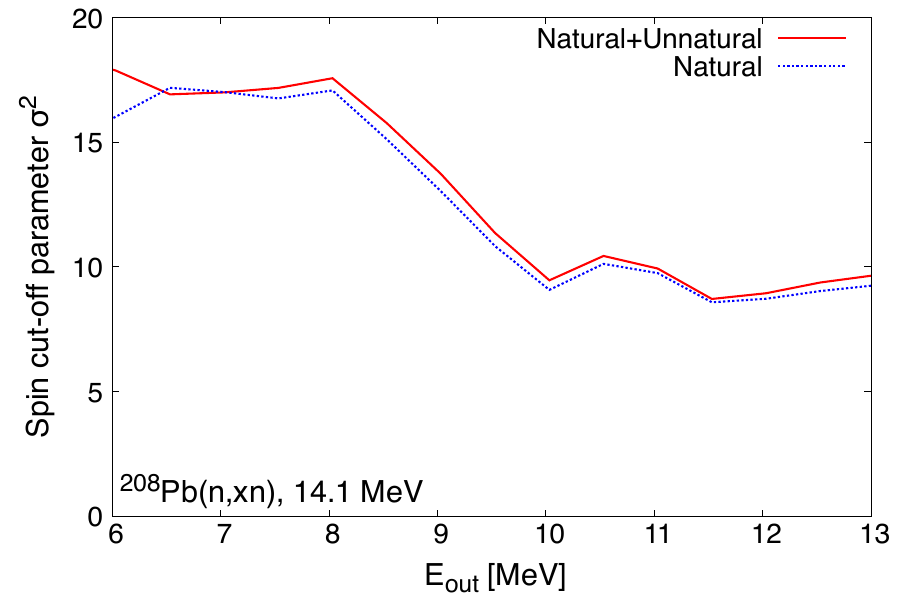}
\caption{
The energy dependence of the spin cut off parameter $\sigma^2$ in Eq.~(\ref{eq:spin dis analytic}) obtained from the correlation between the cut--off parameter and the average spin. The dotted line shows the result considering only natural parity transitions in Eq.~(\ref{eq:spin dis}), while the solid line shows the result considering both natural and unnatural parity transitions.
}
\label{fig:spin sigma}
\end{figure}

\begin{figure}
\includegraphics[width=0.9\linewidth]{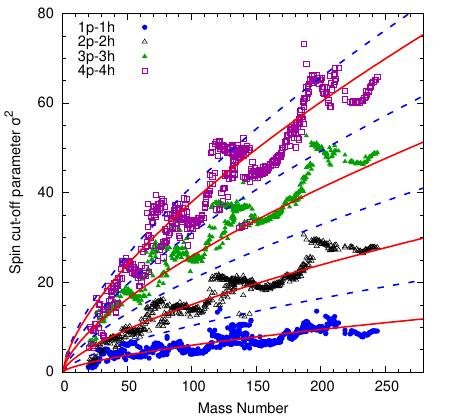}
\caption{
The calculated spin distributions for each of particle-hole configurations expressed in terms of the spin cut-off parameter. The solid and dashed curves are systematics of the parameters as a function of mass number (see text).}
\label{fig:spin coh}
\end{figure}

Figure~\ref{fig:spin distribution} shows the result of Eq.~(\ref{eq:spin dis}) for different outgoing neutron energies from 8 MeV to 11 MeV, where the
compound nucleus decay is negligible. The energy dependence of the spin distribution in Fig.~\ref{fig:spin distribution} reflects the partial contribution from various $J^{\pi}$ states shown in Fig.~\ref{fig:ddw contr}. The large population of the spin $J=3$ for $E_{\mathrm{out}}=11$ MeV in Fig.~\ref{fig:spin distribution} is due to the significant contribution from excitation to the $3^{-}_1(2.615\ \mathrm{MeV})$ state, as seen in Fig.~\ref{fig:ddw contr}. A small bump at $J=5$ reflects the contribution from excitation to the
$5^{-}_1(3.198\ \mathrm{MeV})$ state. For the $E_{\mathrm{out}}=10$ MeV neutron, excitation to the $2^{+}_1(4.086\ \mathrm{MeV})$ state dominates the spin distribution. As the outgoing neutron energy decreases, the excitation to continuum states becomes more prominent, and the spin distribution broadens to higher $J$. The spread of the spin distribution is quantitatively characterized by the value of the spin cut--off parameter $\sigma^2$. The spin distribution of the 1p--1h state derived from Eq.~(\ref{eq:spin dis}) can be fitted by an analytical Gaussian formula \cite{Gilbert:1965vs},
\begin{equation}
\label{eq:spin dis analytic}
    R(J)=\frac{2J+1}{2\sigma^{2}}\exp\left\{-\frac{(J+\frac{1}{2})^2}{2\sigma^2}\right\},
\end{equation}
where $\sigma^2$ is the spin cut--off parameter, and the distribution is normalized as $\sum_{J}R(J)\sim 1$. The spin average is calculated from $\average{J}=\sum_{J}JR(J)$. The spin average and $\sigma$ in Eq.~(\ref{eq:spin dis analytic}) are strongly correlated, and a linear correlation is found numerically: $\average{J}=1.25\sigma-0.52$. This correlation along with the average spin calculated from Eq.~(\ref{eq:spin dis}) is used to estimate the spin cut--off parameters. The results are shown in Fig.~\ref{fig:spin sigma} for various outgoing neutron energies. The dotted line behaves similarly to the solid line due to the small contribution from unnatural parity transitions for $E_{\mathrm{out}}>6$ MeV as seen in Fig.~\ref{fig:ddw contr}. The unnatural parity transition may have an impact on $\sigma^2$ around $E_{\mathrm{out}}=5$ MeV, where the one-step pre-compound reaction becomes comparable to the compound nucleus decay. The increase in $\sigma^2$ as $E_{\mathrm{out}}$ decreases from 10 MeV to 8 MeV in Fig.~\ref{fig:spin sigma} reflects the broadening of the spin distribution as shown in Fig.~\ref{fig:spin distribution}.\\
\indent One of the simplest approaches to evaluate the particle-hole state density is a combinatorial calculation for the single-particle states generated in a one-body potential. We performed the combinatorial calculations for the single-particle spectra in the folding Yukawa potential for various stable isotopes from $Z=10$ to 95, $A=20$ to 244. The axial-symmetric deformation is assumed, and the deformation parameters are taken from the Finite Range Droplet Model (FRDM~\cite{Moller1995,Moller2016}). Since we noticed that the calculated spin-distributions for the particle-hole state densities are relatively insensitive to the excitation energy, we averaged in the excitation energy range 0--50~MeV. The calculated spin distributions are expressed in terms of the spin cut-off parameter and shown in Fig.~\ref{fig:spin coh}. It is known that the spin cut-off parameter is roughly estimated to be $\sigma^2(n) = 0.24 n A^{2/3}$~\cite{Gruppelaar1983}, where $n$ is the number of particles and holes. Although this estimation, shown by the dashed curves in Fig.~\ref{fig:spin coh}, reasonably fits the combinatorial calculation, the fitting can be improved by $\sigma^2(n) = 0.11 (n^2 A)^{2/3}$. This improved fitting provides $\sigma^2(2) = 9.73$ for the 1p-1h excitation of $^{208}\mathrm{Pb}$, which is similar to the values for the pre-equilibrium process in Fig.~\ref{fig:spin sigma}.


\section{Conclusion}\label{sec:Conclusion}


We extended our non-iterative FAM-QRPA method by combining it with DWBA for describing both the direct and one-step pre-equilibrium processes in a consistent manner. This method was applied to the neutron-induced inelastic scattering on $^{208}$Pb. The Skyrme force was used as the interaction between the projectile neutron and the nucleons inside the target nucleus. We demonstrated that the calculated differential cross section for excitations to low-lying states reproduced well the experimental data. We also showed that the two-body and three-body terms in the Skyrme force interfere destructively as reported in Ref.~\cite{Davies:1974ell}. The calculated double differential cross section in the energy regions relevant to the direct and preequilibrium processes shows good agreement with experimental data especially at backward scattering angles. The calculations at forward scattering angles somewhat underestimate the experimental data, which might be improved by including the neglected terms in the Skyrme force used for neutron scatterings. We investigated the spin distribution of populated states in the residual nucleus by summing the partial contributions from individual $J^\Pi$ transition. The values of the spin cut-off parameter seem to be similar to that of the 1p-1h excitation estimated by the combinatorial calculation for the single-particle states.

\begin{acknowledgments}
This work was carried out under the
auspices of the National Nuclear Security Administration of the
U.S. Department of Energy at Los Alamos National Laboratory under
Contract No.~89233218CNA000001. This work was partially supported by the Office of Defense Nuclear Nonproliferation Research \& Development (DNN R\&D), National Nuclear Security Administration,
U.S. Department of Energy. H.S. gratefully acknowledges support by the Advanced Simulation and Computing (ASC) program. 
\end{acknowledgments}
\appendix
\section{$\lambda$ coefficients used for unnatural parity transitions.}
\label{sec:lambda unnatural parity}
In the spherical coordinate $(r,\theta,\phi)$, the contribution from unnatural parity transitions to
Eq.~(\ref{eq:spin expansion}) is calculated from $\lambda$ coefficients in Eq.~(\ref{eq:lambdams}). The $z$-component in Eq.~(\ref{eq:lambdams}) is calculated from the relation of $\sigma_{z}\chi_{\pm1/2}(\sigma)=\pm\chi_{\pm1/2}(\sigma)$,

\begin{eqnarray}
\label{eq:lambdaz}
    \lambda^{z}_{LM}(r)&=&\int\mathrm{d}\Omega\sum_{m}Y_{LM}^{*}(\theta,\phi)\chi^{(-)}_{m\Sigma_{\alpha}}(\vec{r},\vec{k}_{\alpha})^{*}\sigma_{z}\chi^{(+)}_{m\Sigma_{\beta}}(\vec{r},\vec{k}_{\beta})\nonumber\\
    &=&\sum_{J_{\alpha L_{\alpha}}}\sum_{J_{\beta}L_{\beta}}\sum_{m}(-1)^{\Sigma_{\alpha}-2m+\frac{1}{2}}\sqrt{\frac{(2L_{\alpha}+1)(2L_{\beta}+1)}{4\pi(2L+1)}}
    \nonumber\\
    &\times&C^{L_{\alpha},\frac{1}{2},J_{\alpha}}_{M_{\alpha},-\Sigma_{\alpha},M_{\alpha}-\Sigma_{\alpha}}C^{L_{\alpha},\frac{1}{2},J_{\alpha}}_{M_{\alpha}-\Sigma_{\alpha}+m,-m,M_{\alpha}-\Sigma_{\alpha}}\nonumber\\
    &\times&C^{L_{\beta},\frac{1}{2},J_{\beta}}_{0,\Sigma_{\beta},\Sigma_{\beta}}C^{L_{\beta},\frac{1}{2},J_{\beta}}_{\Sigma_{\beta}-m,m,\Sigma_{\beta}}\nonumber\\
    &\times&C^{L_{\alpha},L_{\beta},L}_{M_{\alpha}-\Sigma_{\alpha}+m,\Sigma_{\beta}-m,M}C^{L_{\alpha},L_{\beta},L}_{0,0,0}\nonumber\\
    &\times&\frac{(4\pi)^2}{k_{\alpha}k_{\beta}r^{2}} \chi_{L_{\alpha}J_{\alpha}}^{(+)}(k_{\alpha},r)
    \chi_{L_{\beta}J_{\beta}}^{(+)}(k_{\beta},r)\nonumber\\
    &\times&i^{L_{\alpha}+L_{\beta}}Y^{*}_{L_{\alpha}M_{\alpha}}(-\hat{k}_{\alpha})|_{\phi_{k_\alpha}=0}\sqrt{\frac{2L_{\beta}+1}{4\pi}},
\end{eqnarray}
where $M_{\alpha}=M+\Sigma_{\alpha}-\Sigma_{\beta}$. From the relations $\sigma_{\pm}\chi_{\pm1/2}(\sigma)=0$ and $\sigma_{\pm}\chi_{\mp1/2}(\sigma)=2\chi_{\pm1/2}(\sigma)$, the coefficient $\lambda^{\pm}$ in Eq.~(\ref{eq:lambdams}) is given by
\begin{eqnarray}
\label{eq:lambdaz}
    \lambda^{\pm}_{LM}(r)&=&\int\mathrm{d}\Omega\sum_{m}Y_{LM}^{*}(\theta,\phi)\chi^{(-)}_{m\Sigma_{\alpha}}(\vec{r},\vec{k}_{\alpha})^{*}\sigma_{\pm}\chi^{(+)}_{m\Sigma_{\beta}}(\vec{r},\vec{k}_{\beta})\nonumber\\
    &=&\sum_{J_{\alpha L_{\alpha}}}\sum_{J_{\beta}L_{\beta}}2(-1)^{\Sigma_{\alpha}\mp\frac{1}{2}}\sqrt{\frac{(2L_{\alpha}+1)(2L_{\beta}+1)}{4\pi(2L+1)}}
    \nonumber\\
    &\times&C^{L_{\alpha},\frac{1}{2},J_{\alpha}}_{M_{\alpha},-\Sigma_{\alpha},M_{\alpha}-\Sigma_{\alpha}}C^{L_{\alpha},\frac{1}{2},J_{\alpha}}_{M_{\alpha}-\Sigma_{\alpha}\pm\frac{1}{2},\mp\frac{1}{2},M_{\alpha}-\Sigma_{\alpha}}\nonumber\\
    &\times&C^{L_{\beta},\frac{1}{2},J_{\beta}}_{0,\Sigma_{\beta},\Sigma_{\beta}}C^{L_{\beta},\frac{1}{2},J_{\beta}}_{\Sigma_{\beta}\pm\frac{1}{2},\mp\frac{1}{2},\Sigma_{\beta}}\nonumber\\
    &\times&C^{L_{\alpha},L_{\beta},L}_{M_{\alpha}-\Sigma_{\alpha}\pm\frac{1}{2},\Sigma_{\beta}\pm\frac{1}{2},M}C^{L_{\alpha},L_{\beta},L}_{0,0,0}\nonumber\\
    &\times&\frac{(4\pi)^2}{k_{\alpha}k_{\beta}r^{2}} \chi_{L_{\alpha}J_{\alpha}}^{(+)}(k_{\alpha},r)
    \chi_{L_{\beta}J_{\beta}}^{(+)}(k_{\beta},r)\nonumber\\
    &\times&i^{L_{\alpha}+L_{\beta}}Y^{*}_{L_{\alpha}M_{\alpha}}(-\hat{k}_{\alpha})|_{\phi_{k_\alpha}=0}\sqrt{\frac{2L_{\beta}+1}{4\pi}},
\end{eqnarray}
where $M_{\alpha}=M+\Sigma_{\alpha}-\Sigma_{\beta}\mp1$.

\bibliography{ref}

\begin{thebibliography}{74}%
\makeatletter
\providecommand \@ifxundefined [1]{%
 \@ifx{#1\undefined}
}%
\providecommand \@ifnum [1]{%
 \ifnum #1\expandafter \@firstoftwo
 \else \expandafter \@secondoftwo
 \fi
}%
\providecommand \@ifx [1]{%
 \ifx #1\expandafter \@firstoftwo
 \else \expandafter \@secondoftwo
 \fi
}%
\providecommand \natexlab [1]{#1}%
\providecommand \enquote  [1]{``#1''}%
\providecommand \bibnamefont  [1]{#1}%
\providecommand \bibfnamefont [1]{#1}%
\providecommand \citenamefont [1]{#1}%
\providecommand \href@noop [0]{\@secondoftwo}%
\providecommand \href [0]{\begingroup \@sanitize@url \@href}%
\providecommand \@href[1]{\@@startlink{#1}\@@href}%
\providecommand \@@href[1]{\endgroup#1\@@endlink}%
\providecommand \@sanitize@url [0]{\catcode `\\12\catcode `\$12\catcode
  `\&12\catcode `\#12\catcode `\^12\catcode `\_12\catcode `\%12\relax}%
\providecommand \@@startlink[1]{}%
\providecommand \@@endlink[0]{}%
\providecommand \url  [0]{\begingroup\@sanitize@url \@url }%
\providecommand \@url [1]{\endgroup\@href {#1}{\urlprefix }}%
\providecommand \urlprefix  [0]{URL }%
\providecommand \Eprint [0]{\href }%
\providecommand \doibase [0]{https://doi.org/}%
\providecommand \selectlanguage [0]{\@gobble}%
\providecommand \bibinfo  [0]{\@secondoftwo}%
\providecommand \bibfield  [0]{\@secondoftwo}%
\providecommand \translation [1]{[#1]}%
\providecommand \BibitemOpen [0]{}%
\providecommand \bibitemStop [0]{}%
\providecommand \bibitemNoStop [0]{.\EOS\space}%
\providecommand \EOS [0]{\spacefactor3000\relax}%
\providecommand \BibitemShut  [1]{\csname bibitem#1\endcsname}%
\let\auto@bib@innerbib\@empty
\bibitem [{\citenamefont {Burbidge}\ \emph {et~al.}(1957)\citenamefont
  {Burbidge}, \citenamefont {Burbidge}, \citenamefont {Fowler},\ and\
  \citenamefont {Hoyle}}]{Burbidge:1957vc}%
  \BibitemOpen
  \bibfield  {author} {\bibinfo {author} {\bibfnamefont {M.~E.}\ \bibnamefont
  {Burbidge}}, \bibinfo {author} {\bibfnamefont {G.~R.}\ \bibnamefont
  {Burbidge}}, \bibinfo {author} {\bibfnamefont {W.~A.}\ \bibnamefont
  {Fowler}},\ and\ \bibinfo {author} {\bibfnamefont {F.}~\bibnamefont
  {Hoyle}},\ }\href {https://doi.org/10.1103/RevModPhys.29.547} {\bibfield
  {journal} {\bibinfo  {journal} {Rev. Mod. Phys.}\ }\textbf {\bibinfo {volume}
  {29}},\ \bibinfo {pages} {547} (\bibinfo {year} {1957})}\BibitemShut
  {NoStop}%
\bibitem [{\citenamefont {Arnould}\ \emph {et~al.}(2007)\citenamefont
  {Arnould}, \citenamefont {Goriely},\ and\ \citenamefont
  {Takahashi}}]{ARNOULD200797}%
  \BibitemOpen
  \bibfield  {author} {\bibinfo {author} {\bibfnamefont {M.}~\bibnamefont
  {Arnould}}, \bibinfo {author} {\bibfnamefont {S.}~\bibnamefont {Goriely}},\
  and\ \bibinfo {author} {\bibfnamefont {K.}~\bibnamefont {Takahashi}},\ }\href
  {https://doi.org/https://doi.org/10.1016/j.physrep.2007.06.002} {\bibfield
  {journal} {\bibinfo  {journal} {Physics Reports}\ }\textbf {\bibinfo {volume}
  {450}},\ \bibinfo {pages} {97} (\bibinfo {year} {2007})}\BibitemShut
  {NoStop}%
\bibitem [{\citenamefont {Mumpower}\ \emph {et~al.}(2016)\citenamefont
  {Mumpower}, \citenamefont {Surman}, \citenamefont {McLaughlin},\ and\
  \citenamefont {Aprahamian}}]{Mumpower:2015ova}%
  \BibitemOpen
  \bibfield  {author} {\bibinfo {author} {\bibfnamefont {M.~R.}\ \bibnamefont
  {Mumpower}}, \bibinfo {author} {\bibfnamefont {R.}~\bibnamefont {Surman}},
  \bibinfo {author} {\bibfnamefont {G.~C.}\ \bibnamefont {McLaughlin}},\ and\
  \bibinfo {author} {\bibfnamefont {A.}~\bibnamefont {Aprahamian}},\ }\href
  {https://doi.org/10.1016/j.ppnp.2015.09.001} {\bibfield  {journal} {\bibinfo
  {journal} {Prog. Part. Nucl. Phys.}\ }\textbf {\bibinfo {volume} {86}},\
  \bibinfo {pages} {86} (\bibinfo {year} {2016})}\BibitemShut {NoStop}%
\bibitem [{\citenamefont {Kajino}\ \emph {et~al.}(2019)\citenamefont {Kajino},
  \citenamefont {Aoki}, \citenamefont {Balantekin}, \citenamefont {Diehl},
  \citenamefont {Famiano},\ and\ \citenamefont {Mathews}}]{Kajino:2019abv}%
  \BibitemOpen
  \bibfield  {author} {\bibinfo {author} {\bibfnamefont {T.}~\bibnamefont
  {Kajino}}, \bibinfo {author} {\bibfnamefont {W.}~\bibnamefont {Aoki}},
  \bibinfo {author} {\bibfnamefont {A.~B.}\ \bibnamefont {Balantekin}},
  \bibinfo {author} {\bibfnamefont {R.}~\bibnamefont {Diehl}}, \bibinfo
  {author} {\bibfnamefont {M.~A.}\ \bibnamefont {Famiano}},\ and\ \bibinfo
  {author} {\bibfnamefont {G.~J.}\ \bibnamefont {Mathews}},\ }\href
  {https://doi.org/10.1016/j.ppnp.2019.02.008} {\bibfield  {journal} {\bibinfo
  {journal} {Prog. Part. Nucl. Phys.}\ }\textbf {\bibinfo {volume} {107}},\
  \bibinfo {pages} {109} (\bibinfo {year} {2019})}\BibitemShut {NoStop}%
\bibitem [{\citenamefont {Hayes}\ \emph {et~al.}(2020)\citenamefont {Hayes}
  \emph {et~al.}}]{Hayes:2020cji}%
  \BibitemOpen
  \bibfield  {author} {\bibinfo {author} {\bibfnamefont {A.~C.}\ \bibnamefont
  {Hayes}} \emph {et~al.},\ }\href {https://doi.org/10.1038/s41567-020-0790-3}
  {\bibfield  {journal} {\bibinfo  {journal} {Nature Phys.}\ }\textbf {\bibinfo
  {volume} {16}},\ \bibinfo {pages} {432} (\bibinfo {year} {2020})}\BibitemShut
  {NoStop}%
\bibitem [{\citenamefont {Neudecker}\ \emph {et~al.}(2021)\citenamefont
  {Neudecker}, \citenamefont {Cabellos}, \citenamefont {Clark}, \citenamefont
  {Haeck}, \citenamefont {Capote}, \citenamefont {Trkov}, \citenamefont
  {White},\ and\ \citenamefont {Rising}}]{NEUDECKER2021108345}%
  \BibitemOpen
  \bibfield  {author} {\bibinfo {author} {\bibfnamefont {D.}~\bibnamefont
  {Neudecker}}, \bibinfo {author} {\bibfnamefont {O.}~\bibnamefont {Cabellos}},
  \bibinfo {author} {\bibfnamefont {A.~R.}\ \bibnamefont {Clark}}, \bibinfo
  {author} {\bibfnamefont {W.}~\bibnamefont {Haeck}}, \bibinfo {author}
  {\bibfnamefont {R.}~\bibnamefont {Capote}}, \bibinfo {author} {\bibfnamefont
  {A.}~\bibnamefont {Trkov}}, \bibinfo {author} {\bibfnamefont {M.~C.}\
  \bibnamefont {White}},\ and\ \bibinfo {author} {\bibfnamefont {M.~E.}\
  \bibnamefont {Rising}},\ }\href
  {https://doi.org/https://doi.org/10.1016/j.anucene.2021.108345} {\bibfield
  {journal} {\bibinfo  {journal} {Annals of Nuclear Energy}\ }\textbf {\bibinfo
  {volume} {159}},\ \bibinfo {pages} {108345} (\bibinfo {year}
  {2021})}\BibitemShut {NoStop}%
\bibitem [{\citenamefont {Gadioli}\ and\ \citenamefont
  {Hodgson}(1992)}]{gadioli1992pre}%
  \BibitemOpen
  \bibfield  {author} {\bibinfo {author} {\bibfnamefont {E.}~\bibnamefont
  {Gadioli}}\ and\ \bibinfo {author} {\bibfnamefont {P.}~\bibnamefont
  {Hodgson}},\ }\href {https://books.google.com/books?id=lJp9AAAAIAAJ} {\emph
  {\bibinfo {title} {Pre-equilibrium Nuclear Reactions}}},\ Oxford science
  publications\ (\bibinfo  {publisher} {Clarendon Press},\ \bibinfo {year}
  {1992})\BibitemShut {NoStop}%
\bibitem [{\citenamefont {Hauser}\ and\ \citenamefont
  {Feshbach}(1952)}]{Hauser:1952zz}%
  \BibitemOpen
  \bibfield  {author} {\bibinfo {author} {\bibfnamefont {W.}~\bibnamefont
  {Hauser}}\ and\ \bibinfo {author} {\bibfnamefont {H.}~\bibnamefont
  {Feshbach}},\ }\href {https://doi.org/10.1103/PhysRev.87.366} {\bibfield
  {journal} {\bibinfo  {journal} {Phys. Rev.}\ }\textbf {\bibinfo {volume}
  {87}},\ \bibinfo {pages} {366} (\bibinfo {year} {1952})}\BibitemShut
  {NoStop}%
\bibitem [{\citenamefont {Glendenning}(2004)}]{glendenning2004direct}%
  \BibitemOpen
  \bibfield  {author} {\bibinfo {author} {\bibfnamefont {N.}~\bibnamefont
  {Glendenning}},\ }\href {https://books.google.com/books?id=UrxpDQAAQBAJ}
  {\emph {\bibinfo {title} {Direct Nuclear Reactions}}}\ (\bibinfo  {publisher}
  {World Scientific},\ \bibinfo {year} {2004})\BibitemShut {NoStop}%
\bibitem [{\citenamefont {Griffin}(1966)}]{GriffinPRL1966}%
  \BibitemOpen
  \bibfield  {author} {\bibinfo {author} {\bibfnamefont {J.~J.}\ \bibnamefont
  {Griffin}},\ }\href {https://doi.org/10.1103/PhysRevLett.17.478} {\bibfield
  {journal} {\bibinfo  {journal} {Phys. Rev. Lett.}\ }\textbf {\bibinfo
  {volume} {17}},\ \bibinfo {pages} {478} (\bibinfo {year} {1966})}\BibitemShut
  {NoStop}%
\bibitem [{\citenamefont {Blann}(1971)}]{BlannPRL1971}%
  \BibitemOpen
  \bibfield  {author} {\bibinfo {author} {\bibfnamefont {M.}~\bibnamefont
  {Blann}},\ }\href {https://doi.org/10.1103/PhysRevLett.27.337} {\bibfield
  {journal} {\bibinfo  {journal} {Phys. Rev. Lett.}\ }\textbf {\bibinfo
  {volume} {27}},\ \bibinfo {pages} {337} (\bibinfo {year} {1971})}\BibitemShut
  {NoStop}%
\bibitem [{\citenamefont {Mantzouranis}\ \emph {et~al.}(1976)\citenamefont
  {Mantzouranis}, \citenamefont {Weidenm{\"u}ller},\ and\ \citenamefont
  {Agassi}}]{mantzouranis1976generalized}%
  \BibitemOpen
  \bibfield  {author} {\bibinfo {author} {\bibfnamefont {G.}~\bibnamefont
  {Mantzouranis}}, \bibinfo {author} {\bibfnamefont {H.}~\bibnamefont
  {Weidenm{\"u}ller}},\ and\ \bibinfo {author} {\bibfnamefont {D.}~\bibnamefont
  {Agassi}},\ }\href {https://link.springer.com/article/10.1007/BF01437709}
  {\bibfield  {journal} {\bibinfo  {journal} {Zeitschrift f{\"u}r Physik A
  Atoms and Nuclei}\ }\textbf {\bibinfo {volume} {276}},\ \bibinfo {pages}
  {145} (\bibinfo {year} {1976})}\BibitemShut {NoStop}%
\bibitem [{\citenamefont {Blann}\ \emph {et~al.}(1984)\citenamefont {Blann},
  \citenamefont {Scobel},\ and\ \citenamefont {Plechaty}}]{BlannPRC1984}%
  \BibitemOpen
  \bibfield  {author} {\bibinfo {author} {\bibfnamefont {M.}~\bibnamefont
  {Blann}}, \bibinfo {author} {\bibfnamefont {W.}~\bibnamefont {Scobel}},\ and\
  \bibinfo {author} {\bibfnamefont {E.}~\bibnamefont {Plechaty}},\ }\href
  {https://doi.org/10.1103/PhysRevC.30.1493} {\bibfield  {journal} {\bibinfo
  {journal} {Phys. Rev. C}\ }\textbf {\bibinfo {volume} {30}},\ \bibinfo
  {pages} {1493} (\bibinfo {year} {1984})}\BibitemShut {NoStop}%
\bibitem [{\citenamefont {Kawano}\ \emph {et~al.}(2006)\citenamefont {Kawano},
  \citenamefont {Talou},\ and\ \citenamefont {Chadwick}}]{KAWANO2006774}%
  \BibitemOpen
  \bibfield  {author} {\bibinfo {author} {\bibfnamefont {T.}~\bibnamefont
  {Kawano}}, \bibinfo {author} {\bibfnamefont {P.}~\bibnamefont {Talou}},\ and\
  \bibinfo {author} {\bibfnamefont {M.~B.}\ \bibnamefont {Chadwick}},\ }\href
  {https://doi.org/https://doi.org/10.1016/j.nima.2006.02.053} {\bibfield
  {journal} {\bibinfo  {journal} {Nuclear Instruments and Methods in Physics
  Research Section A: Accelerators, Spectrometers, Detectors and Associated
  Equipment}\ }\textbf {\bibinfo {volume} {562}},\ \bibinfo {pages} {774}
  (\bibinfo {year} {2006})},\ \bibinfo {note} {proceedings of the 7th
  International Conference on Accelerator Applications}\BibitemShut {NoStop}%
\bibitem [{\citenamefont {Mumpower}\ \emph {et~al.}(2023)\citenamefont
  {Mumpower}, \citenamefont {Neudecker}, \citenamefont {Sasaki}, \citenamefont
  {Kawano}, \citenamefont {Lovell}, \citenamefont {Herman}, \citenamefont
  {Stetcu},\ and\ \citenamefont {Dupuis}}]{MumpowerPRC2023}%
  \BibitemOpen
  \bibfield  {author} {\bibinfo {author} {\bibfnamefont {M.~R.}\ \bibnamefont
  {Mumpower}}, \bibinfo {author} {\bibfnamefont {D.}~\bibnamefont {Neudecker}},
  \bibinfo {author} {\bibfnamefont {H.}~\bibnamefont {Sasaki}}, \bibinfo
  {author} {\bibfnamefont {T.}~\bibnamefont {Kawano}}, \bibinfo {author}
  {\bibfnamefont {A.~E.}\ \bibnamefont {Lovell}}, \bibinfo {author}
  {\bibfnamefont {M.~W.}\ \bibnamefont {Herman}}, \bibinfo {author}
  {\bibfnamefont {I.}~\bibnamefont {Stetcu}},\ and\ \bibinfo {author}
  {\bibfnamefont {M.}~\bibnamefont {Dupuis}},\ }\href
  {https://doi.org/10.1103/PhysRevC.107.034606} {\bibfield  {journal} {\bibinfo
   {journal} {Phys. Rev. C}\ }\textbf {\bibinfo {volume} {107}},\ \bibinfo
  {pages} {034606} (\bibinfo {year} {2023})}\BibitemShut {NoStop}%
\bibitem [{\citenamefont {Feshbach}\ \emph {et~al.}(1980)\citenamefont
  {Feshbach}, \citenamefont {Kerman},\ and\ \citenamefont
  {Koonin}}]{FESHBACH1980429}%
  \BibitemOpen
  \bibfield  {author} {\bibinfo {author} {\bibfnamefont {H.}~\bibnamefont
  {Feshbach}}, \bibinfo {author} {\bibfnamefont {A.}~\bibnamefont {Kerman}},\
  and\ \bibinfo {author} {\bibfnamefont {S.}~\bibnamefont {Koonin}},\ }\href
  {https://doi.org/https://doi.org/10.1016/0003-4916(80)90140-2} {\bibfield
  {journal} {\bibinfo  {journal} {Annals of Physics}\ }\textbf {\bibinfo
  {volume} {125}},\ \bibinfo {pages} {429} (\bibinfo {year}
  {1980})}\BibitemShut {NoStop}%
\bibitem [{\citenamefont {Tamura}\ \emph {et~al.}(1982)\citenamefont {Tamura},
  \citenamefont {Udagawa},\ and\ \citenamefont
  {Lenske}}]{Tamura-Udagawa-Lenske}%
  \BibitemOpen
  \bibfield  {author} {\bibinfo {author} {\bibfnamefont {T.}~\bibnamefont
  {Tamura}}, \bibinfo {author} {\bibfnamefont {T.}~\bibnamefont {Udagawa}},\
  and\ \bibinfo {author} {\bibfnamefont {H.}~\bibnamefont {Lenske}},\ }\href
  {https://doi.org/10.1103/PhysRevC.26.379} {\bibfield  {journal} {\bibinfo
  {journal} {Phys. Rev. C}\ }\textbf {\bibinfo {volume} {26}},\ \bibinfo
  {pages} {379} (\bibinfo {year} {1982})}\BibitemShut {NoStop}%
\bibitem [{\citenamefont {Nishioka}\ \emph {et~al.}(1989)\citenamefont
  {Nishioka}, \citenamefont {Weidenmüller},\ and\ \citenamefont
  {Yoshida}}]{NISHIOKA1989195}%
  \BibitemOpen
  \bibfield  {author} {\bibinfo {author} {\bibfnamefont {H.}~\bibnamefont
  {Nishioka}}, \bibinfo {author} {\bibfnamefont {H.}~\bibnamefont
  {Weidenmüller}},\ and\ \bibinfo {author} {\bibfnamefont {S.}~\bibnamefont
  {Yoshida}},\ }\href
  {https://doi.org/https://doi.org/10.1016/0003-4916(89)90358-8} {\bibfield
  {journal} {\bibinfo  {journal} {Annals of Physics}\ }\textbf {\bibinfo
  {volume} {193}},\ \bibinfo {pages} {195} (\bibinfo {year}
  {1989})}\BibitemShut {NoStop}%
\bibitem [{\citenamefont {Koning}\ and\ \citenamefont
  {Akkermans}(1991)}]{KONING1991216}%
  \BibitemOpen
  \bibfield  {author} {\bibinfo {author} {\bibfnamefont {A.}~\bibnamefont
  {Koning}}\ and\ \bibinfo {author} {\bibfnamefont {J.}~\bibnamefont
  {Akkermans}},\ }\href
  {https://doi.org/https://doi.org/10.1016/0003-4916(91)90345-9} {\bibfield
  {journal} {\bibinfo  {journal} {Annals of Physics}\ }\textbf {\bibinfo
  {volume} {208}},\ \bibinfo {pages} {216} (\bibinfo {year}
  {1991})}\BibitemShut {NoStop}%
\bibitem [{\citenamefont {Kawano}\ and\ \citenamefont
  {Yoshida}(2001)}]{KawanoPRC2002}%
  \BibitemOpen
  \bibfield  {author} {\bibinfo {author} {\bibfnamefont {T.}~\bibnamefont
  {Kawano}}\ and\ \bibinfo {author} {\bibfnamefont {S.}~\bibnamefont
  {Yoshida}},\ }\href {https://doi.org/10.1103/PhysRevC.64.024603} {\bibfield
  {journal} {\bibinfo  {journal} {Phys. Rev. C}\ }\textbf {\bibinfo {volume}
  {64}},\ \bibinfo {pages} {024603} (\bibinfo {year} {2001})}\BibitemShut
  {NoStop}%
\bibitem [{\citenamefont {Chadwick}\ and\ \citenamefont
  {Young}(1993)}]{ChadwickPRC1993}%
  \BibitemOpen
  \bibfield  {author} {\bibinfo {author} {\bibfnamefont {M.~B.}\ \bibnamefont
  {Chadwick}}\ and\ \bibinfo {author} {\bibfnamefont {P.~G.}\ \bibnamefont
  {Young}},\ }\href {https://doi.org/10.1103/PhysRevC.47.2255} {\bibfield
  {journal} {\bibinfo  {journal} {Phys. Rev. C}\ }\textbf {\bibinfo {volume}
  {47}},\ \bibinfo {pages} {2255} (\bibinfo {year} {1993})}\BibitemShut
  {NoStop}%
\bibitem [{\citenamefont {Ring}\ and\ \citenamefont
  {Schuck}(2004)}]{ring2004nuclear}%
  \BibitemOpen
  \bibfield  {author} {\bibinfo {author} {\bibfnamefont {P.}~\bibnamefont
  {Ring}}\ and\ \bibinfo {author} {\bibfnamefont {P.}~\bibnamefont {Schuck}},\
  }\href@noop {} {\emph {\bibinfo {title} {The Nuclear Many-Body Problem}}},\
  Physics and astronomy online library\ (\bibinfo  {publisher} {Springer},\
  \bibinfo {year} {2004})\BibitemShut {NoStop}%
\bibitem [{\citenamefont {Dupuis}\ \emph {et~al.}(2011)\citenamefont {Dupuis},
  \citenamefont {Kawano}, \citenamefont {Delaroche},\ and\ \citenamefont
  {Bauge}}]{Dupuis:2011zza}%
  \BibitemOpen
  \bibfield  {author} {\bibinfo {author} {\bibfnamefont {M.}~\bibnamefont
  {Dupuis}}, \bibinfo {author} {\bibfnamefont {T.}~\bibnamefont {Kawano}},
  \bibinfo {author} {\bibfnamefont {J.~P.}\ \bibnamefont {Delaroche}},\ and\
  \bibinfo {author} {\bibfnamefont {E.}~\bibnamefont {Bauge}},\ }\href
  {https://doi.org/10.1103/PhysRevC.83.014602} {\bibfield  {journal} {\bibinfo
  {journal} {Phys. Rev. C}\ }\textbf {\bibinfo {volume} {83}},\ \bibinfo
  {pages} {014602} (\bibinfo {year} {2011})}\BibitemShut {NoStop}%
\bibitem [{\citenamefont {Dupuis}\ \emph {et~al.}(2015)\citenamefont {Dupuis},
  \citenamefont {Bauge}, \citenamefont {Hilaire}, \citenamefont {Lechaftois},
  \citenamefont {P\'eru}, \citenamefont {Pillet},\ and\ \citenamefont
  {Robin}}]{Dupuis:2015uqy}%
  \BibitemOpen
  \bibfield  {author} {\bibinfo {author} {\bibfnamefont {M.}~\bibnamefont
  {Dupuis}}, \bibinfo {author} {\bibfnamefont {E.}~\bibnamefont {Bauge}},
  \bibinfo {author} {\bibfnamefont {S.}~\bibnamefont {Hilaire}}, \bibinfo
  {author} {\bibfnamefont {F.}~\bibnamefont {Lechaftois}}, \bibinfo {author}
  {\bibfnamefont {S.}~\bibnamefont {P\'eru}}, \bibinfo {author} {\bibfnamefont
  {N.}~\bibnamefont {Pillet}},\ and\ \bibinfo {author} {\bibfnamefont
  {C.}~\bibnamefont {Robin}},\ }\href
  {https://doi.org/10.1140/epja/i2015-15168-x} {\bibfield  {journal} {\bibinfo
  {journal} {Eur. Phys. J. A}\ }\textbf {\bibinfo {volume} {51}},\ \bibinfo
  {pages} {168} (\bibinfo {year} {2015})}\BibitemShut {NoStop}%
\bibitem [{\citenamefont {Dupuis}\ \emph {et~al.}(2017)\citenamefont {Dupuis},
  \citenamefont {Hilaire}, \citenamefont {P{\'e}ru}, \citenamefont {Bauge},
  \citenamefont {Kerveno}, \citenamefont {Dessagne},\ and\ \citenamefont
  {Henning}}]{dupuis2017microscopic}%
  \BibitemOpen
  \bibfield  {author} {\bibinfo {author} {\bibfnamefont {M.}~\bibnamefont
  {Dupuis}}, \bibinfo {author} {\bibfnamefont {S.}~\bibnamefont {Hilaire}},
  \bibinfo {author} {\bibfnamefont {S.}~\bibnamefont {P{\'e}ru}}, \bibinfo
  {author} {\bibfnamefont {E.}~\bibnamefont {Bauge}}, \bibinfo {author}
  {\bibfnamefont {M.}~\bibnamefont {Kerveno}}, \bibinfo {author} {\bibfnamefont
  {P.}~\bibnamefont {Dessagne}},\ and\ \bibinfo {author} {\bibfnamefont
  {G.}~\bibnamefont {Henning}},\ }in\ \href
  {https://www.epj-conferences.org/articles/epjconf/abs/2017/15/epjconf-nd2016_12002/epjconf-nd2016_12002.html}
  {\emph {\bibinfo {booktitle} {EPJ Web of Conferences}}},\ Vol.\ \bibinfo
  {volume} {146}\ (\bibinfo {organization} {EDP Sciences},\ \bibinfo {year}
  {2017})\ p.\ \bibinfo {pages} {12002}\BibitemShut {NoStop}%
\bibitem [{\citenamefont {Dupuis}\ \emph {et~al.}(2019)\citenamefont {Dupuis},
  \citenamefont {Haouat}, \citenamefont {Delaroche}, \citenamefont {Bauge},\
  and\ \citenamefont {Lachkar}}]{Dupuis2019}%
  \BibitemOpen
  \bibfield  {author} {\bibinfo {author} {\bibfnamefont {M.}~\bibnamefont
  {Dupuis}}, \bibinfo {author} {\bibfnamefont {G.}~\bibnamefont {Haouat}},
  \bibinfo {author} {\bibfnamefont {J.-P.}\ \bibnamefont {Delaroche}}, \bibinfo
  {author} {\bibfnamefont {E.}~\bibnamefont {Bauge}},\ and\ \bibinfo {author}
  {\bibfnamefont {J.}~\bibnamefont {Lachkar}},\ }\href
  {https://doi.org/10.1103/PhysRevC.100.044607} {\bibfield  {journal} {\bibinfo
   {journal} {Phys. Rev. C}\ }\textbf {\bibinfo {volume} {100}},\ \bibinfo
  {pages} {044607} (\bibinfo {year} {2019})}\BibitemShut {NoStop}%
\bibitem [{\citenamefont {Nakatsukasa}\ \emph {et~al.}(2007)\citenamefont
  {Nakatsukasa}, \citenamefont {Inakura},\ and\ \citenamefont
  {Yabana}}]{Nakatsukasa:2007qj}%
  \BibitemOpen
  \bibfield  {author} {\bibinfo {author} {\bibfnamefont {T.}~\bibnamefont
  {Nakatsukasa}}, \bibinfo {author} {\bibfnamefont {T.}~\bibnamefont
  {Inakura}},\ and\ \bibinfo {author} {\bibfnamefont {K.}~\bibnamefont
  {Yabana}},\ }\href {https://doi.org/10.1103/PhysRevC.76.024318} {\bibfield
  {journal} {\bibinfo  {journal} {Phys. Rev. C}\ }\textbf {\bibinfo {volume}
  {76}},\ \bibinfo {pages} {024318} (\bibinfo {year} {2007})}\BibitemShut
  {NoStop}%
\bibitem [{\citenamefont {Avogadro}\ and\ \citenamefont
  {Nakatsukasa}(2011)}]{Avogadro:2011gd}%
  \BibitemOpen
  \bibfield  {author} {\bibinfo {author} {\bibfnamefont {P.}~\bibnamefont
  {Avogadro}}\ and\ \bibinfo {author} {\bibfnamefont {T.}~\bibnamefont
  {Nakatsukasa}},\ }\href {https://doi.org/10.1103/PhysRevC.84.014314}
  {\bibfield  {journal} {\bibinfo  {journal} {Phys. Rev. C}\ }\textbf {\bibinfo
  {volume} {84}},\ \bibinfo {pages} {014314} (\bibinfo {year}
  {2011})}\BibitemShut {NoStop}%
\bibitem [{\citenamefont {Inakura}\ \emph {et~al.}(2009)\citenamefont
  {Inakura}, \citenamefont {Nakatsukasa},\ and\ \citenamefont
  {Yabana}}]{Inakura:2009vs}%
  \BibitemOpen
  \bibfield  {author} {\bibinfo {author} {\bibfnamefont {T.}~\bibnamefont
  {Inakura}}, \bibinfo {author} {\bibfnamefont {T.}~\bibnamefont
  {Nakatsukasa}},\ and\ \bibinfo {author} {\bibfnamefont {K.}~\bibnamefont
  {Yabana}},\ }\href {https://doi.org/10.1103/PhysRevC.80.044301} {\bibfield
  {journal} {\bibinfo  {journal} {Phys. Rev. C}\ }\textbf {\bibinfo {volume}
  {80}},\ \bibinfo {pages} {044301} (\bibinfo {year} {2009})}\BibitemShut
  {NoStop}%
\bibitem [{\citenamefont {Stoitsov}\ \emph {et~al.}(2011)\citenamefont
  {Stoitsov}, \citenamefont {Kortelainen}, \citenamefont {Nakatsukasa},
  \citenamefont {Losa},\ and\ \citenamefont {Nazarewicz}}]{Stoitsov:2011zz}%
  \BibitemOpen
  \bibfield  {author} {\bibinfo {author} {\bibfnamefont {M.}~\bibnamefont
  {Stoitsov}}, \bibinfo {author} {\bibfnamefont {M.}~\bibnamefont
  {Kortelainen}}, \bibinfo {author} {\bibfnamefont {T.}~\bibnamefont
  {Nakatsukasa}}, \bibinfo {author} {\bibfnamefont {C.}~\bibnamefont {Losa}},\
  and\ \bibinfo {author} {\bibfnamefont {W.}~\bibnamefont {Nazarewicz}},\
  }\href {https://doi.org/10.1103/PhysRevC.84.041305} {\bibfield  {journal}
  {\bibinfo  {journal} {Phys. Rev. C}\ }\textbf {\bibinfo {volume} {84}},\
  \bibinfo {pages} {041305} (\bibinfo {year} {2011})}\BibitemShut {NoStop}%
\bibitem [{\citenamefont {Hinohara}\ \emph {et~al.}(2013)\citenamefont
  {Hinohara}, \citenamefont {Kortelainen},\ and\ \citenamefont
  {Nazarewicz}}]{Hinohara:2013qda}%
  \BibitemOpen
  \bibfield  {author} {\bibinfo {author} {\bibfnamefont {N.}~\bibnamefont
  {Hinohara}}, \bibinfo {author} {\bibfnamefont {M.}~\bibnamefont
  {Kortelainen}},\ and\ \bibinfo {author} {\bibfnamefont {W.}~\bibnamefont
  {Nazarewicz}},\ }\href {https://doi.org/10.1103/PhysRevC.87.064309}
  {\bibfield  {journal} {\bibinfo  {journal} {Phys. Rev. C}\ }\textbf {\bibinfo
  {volume} {87}},\ \bibinfo {pages} {064309} (\bibinfo {year}
  {2013})}\BibitemShut {NoStop}%
\bibitem [{\citenamefont {Kortelainen}\ \emph {et~al.}(2015)\citenamefont
  {Kortelainen}, \citenamefont {Hinohara},\ and\ \citenamefont
  {Nazarewicz}}]{Kortelainen:2015gxa}%
  \BibitemOpen
  \bibfield  {author} {\bibinfo {author} {\bibfnamefont {M.}~\bibnamefont
  {Kortelainen}}, \bibinfo {author} {\bibfnamefont {N.}~\bibnamefont
  {Hinohara}},\ and\ \bibinfo {author} {\bibfnamefont {W.}~\bibnamefont
  {Nazarewicz}},\ }\href {https://doi.org/10.1103/PhysRevC.92.051302}
  {\bibfield  {journal} {\bibinfo  {journal} {Phys. Rev. C}\ }\textbf {\bibinfo
  {volume} {92}},\ \bibinfo {pages} {051302} (\bibinfo {year}
  {2015})}\BibitemShut {NoStop}%
\bibitem [{\citenamefont {Oishi}\ \emph {et~al.}(2016)\citenamefont {Oishi},
  \citenamefont {Kortelainen},\ and\ \citenamefont {Hinohara}}]{Oishi:2015lph}%
  \BibitemOpen
  \bibfield  {author} {\bibinfo {author} {\bibfnamefont {T.}~\bibnamefont
  {Oishi}}, \bibinfo {author} {\bibfnamefont {M.}~\bibnamefont {Kortelainen}},\
  and\ \bibinfo {author} {\bibfnamefont {N.}~\bibnamefont {Hinohara}},\ }\href
  {https://doi.org/10.1103/PhysRevC.93.034329} {\bibfield  {journal} {\bibinfo
  {journal} {Phys. Rev. C}\ }\textbf {\bibinfo {volume} {93}},\ \bibinfo
  {pages} {034329} (\bibinfo {year} {2016})}\BibitemShut {NoStop}%
\bibitem [{\citenamefont {Sasaki}\ \emph {et~al.}(2022)\citenamefont {Sasaki},
  \citenamefont {Kawano},\ and\ \citenamefont {Stetcu}}]{Sasaki:2022ipn}%
  \BibitemOpen
  \bibfield  {author} {\bibinfo {author} {\bibfnamefont {H.}~\bibnamefont
  {Sasaki}}, \bibinfo {author} {\bibfnamefont {T.}~\bibnamefont {Kawano}},\
  and\ \bibinfo {author} {\bibfnamefont {I.}~\bibnamefont {Stetcu}},\ }\href
  {https://doi.org/10.1103/PhysRevC.105.044311} {\bibfield  {journal} {\bibinfo
   {journal} {Phys. Rev. C}\ }\textbf {\bibinfo {volume} {105}},\ \bibinfo
  {pages} {044311} (\bibinfo {year} {2022})}\BibitemShut {NoStop}%
\bibitem [{\citenamefont {Sasaki}\ \emph {et~al.}(2023)\citenamefont {Sasaki},
  \citenamefont {Kawano},\ and\ \citenamefont {Stetcu}}]{SasakiPRC2023}%
  \BibitemOpen
  \bibfield  {author} {\bibinfo {author} {\bibfnamefont {H.}~\bibnamefont
  {Sasaki}}, \bibinfo {author} {\bibfnamefont {T.}~\bibnamefont {Kawano}},\
  and\ \bibinfo {author} {\bibfnamefont {I.}~\bibnamefont {Stetcu}},\ }\href
  {https://doi.org/10.1103/PhysRevC.107.054312} {\bibfield  {journal} {\bibinfo
   {journal} {Phys. Rev. C}\ }\textbf {\bibinfo {volume} {107}},\ \bibinfo
  {pages} {054312} (\bibinfo {year} {2023})}\BibitemShut {NoStop}%
\bibitem [{\citenamefont {Washiyama}\ \emph {et~al.}(2024)\citenamefont
  {Washiyama}, \citenamefont {Ebata},\ and\ \citenamefont
  {Yoshida}}]{Washiyama:2023jxg}%
  \BibitemOpen
  \bibfield  {author} {\bibinfo {author} {\bibfnamefont {K.}~\bibnamefont
  {Washiyama}}, \bibinfo {author} {\bibfnamefont {S.}~\bibnamefont {Ebata}},\
  and\ \bibinfo {author} {\bibfnamefont {K.}~\bibnamefont {Yoshida}},\ }\href
  {https://doi.org/10.1103/PhysRevC.109.024317} {\bibfield  {journal} {\bibinfo
   {journal} {Phys. Rev. C}\ }\textbf {\bibinfo {volume} {109}},\ \bibinfo
  {pages} {024317} (\bibinfo {year} {2024})}\BibitemShut {NoStop}%
\bibitem [{\citenamefont {Frosini}\ \emph {et~al.}(2024)\citenamefont
  {Frosini}, \citenamefont {Ryssens},\ and\ \citenamefont
  {Sieja}}]{Frosini:2023bym}%
  \BibitemOpen
  \bibfield  {author} {\bibinfo {author} {\bibfnamefont {M.}~\bibnamefont
  {Frosini}}, \bibinfo {author} {\bibfnamefont {W.}~\bibnamefont {Ryssens}},\
  and\ \bibinfo {author} {\bibfnamefont {K.}~\bibnamefont {Sieja}},\ }\href
  {https://doi.org/10.1103/PhysRevC.110.014307} {\bibfield  {journal} {\bibinfo
   {journal} {Phys. Rev. C}\ }\textbf {\bibinfo {volume} {110}},\ \bibinfo
  {pages} {014307} (\bibinfo {year} {2024})}\BibitemShut {NoStop}%
\bibitem [{\citenamefont {Li}\ \emph {et~al.}(2024)\citenamefont {Li},
  \citenamefont {Schunck},\ and\ \citenamefont {Grosskopf}}]{Li:2024ivd}%
  \BibitemOpen
  \bibfield  {author} {\bibinfo {author} {\bibfnamefont {T.}~\bibnamefont
  {Li}}, \bibinfo {author} {\bibfnamefont {N.}~\bibnamefont {Schunck}},\ and\
  \bibinfo {author} {\bibfnamefont {M.}~\bibnamefont {Grosskopf}},\ }\href
  {https://doi.org/10.1103/PhysRevC.110.034317} {\bibfield  {journal} {\bibinfo
   {journal} {Phys. Rev. C}\ }\textbf {\bibinfo {volume} {110}},\ \bibinfo
  {pages} {034317} (\bibinfo {year} {2024})}\BibitemShut {NoStop}%
\bibitem [{\citenamefont {Mustonen}\ \emph {et~al.}(2014)\citenamefont
  {Mustonen}, \citenamefont {Shafer}, \citenamefont {Zenginerler},\ and\
  \citenamefont {Engel}}]{Mustonen:2014bya}%
  \BibitemOpen
  \bibfield  {author} {\bibinfo {author} {\bibfnamefont {M.~T.}\ \bibnamefont
  {Mustonen}}, \bibinfo {author} {\bibfnamefont {T.}~\bibnamefont {Shafer}},
  \bibinfo {author} {\bibfnamefont {Z.}~\bibnamefont {Zenginerler}},\ and\
  \bibinfo {author} {\bibfnamefont {J.}~\bibnamefont {Engel}},\ }\href
  {https://doi.org/10.1103/PhysRevC.90.024308} {\bibfield  {journal} {\bibinfo
  {journal} {Phys. Rev. C}\ }\textbf {\bibinfo {volume} {90}},\ \bibinfo
  {pages} {024308} (\bibinfo {year} {2014})}\BibitemShut {NoStop}%
\bibitem [{\citenamefont {Shafer}\ \emph {et~al.}(2016)\citenamefont {Shafer},
  \citenamefont {Engel}, \citenamefont {Fr\"ohlich}, \citenamefont
  {McLaughlin}, \citenamefont {Mumpower},\ and\ \citenamefont
  {Surman}}]{Shafer:2016etk}%
  \BibitemOpen
  \bibfield  {author} {\bibinfo {author} {\bibfnamefont {T.}~\bibnamefont
  {Shafer}}, \bibinfo {author} {\bibfnamefont {J.}~\bibnamefont {Engel}},
  \bibinfo {author} {\bibfnamefont {C.}~\bibnamefont {Fr\"ohlich}}, \bibinfo
  {author} {\bibfnamefont {G.~C.}\ \bibnamefont {McLaughlin}}, \bibinfo
  {author} {\bibfnamefont {M.}~\bibnamefont {Mumpower}},\ and\ \bibinfo
  {author} {\bibfnamefont {R.}~\bibnamefont {Surman}},\ }\href
  {https://doi.org/10.1103/PhysRevC.94.055802} {\bibfield  {journal} {\bibinfo
  {journal} {Phys. Rev. C}\ }\textbf {\bibinfo {volume} {94}},\ \bibinfo
  {pages} {055802} (\bibinfo {year} {2016})}\BibitemShut {NoStop}%
\bibitem [{\citenamefont {Ney}\ \emph {et~al.}(2020)\citenamefont {Ney},
  \citenamefont {Engel}, \citenamefont {Li},\ and\ \citenamefont
  {Schunck}}]{Ney:2020mnx}%
  \BibitemOpen
  \bibfield  {author} {\bibinfo {author} {\bibfnamefont {E.~M.}\ \bibnamefont
  {Ney}}, \bibinfo {author} {\bibfnamefont {J.}~\bibnamefont {Engel}}, \bibinfo
  {author} {\bibfnamefont {T.}~\bibnamefont {Li}},\ and\ \bibinfo {author}
  {\bibfnamefont {N.}~\bibnamefont {Schunck}},\ }\href
  {https://doi.org/10.1103/PhysRevC.102.034326} {\bibfield  {journal} {\bibinfo
   {journal} {Phys. Rev. C}\ }\textbf {\bibinfo {volume} {102}},\ \bibinfo
  {pages} {034326} (\bibinfo {year} {2020})}\BibitemShut {NoStop}%
\bibitem [{\citenamefont {Hinohara}\ and\ \citenamefont
  {Engel}(2022)}]{Hinohara:2022uip}%
  \BibitemOpen
  \bibfield  {author} {\bibinfo {author} {\bibfnamefont {N.}~\bibnamefont
  {Hinohara}}\ and\ \bibinfo {author} {\bibfnamefont {J.}~\bibnamefont
  {Engel}},\ }\href {https://doi.org/10.1103/PhysRevC.105.044314} {\bibfield
  {journal} {\bibinfo  {journal} {Phys. Rev. C}\ }\textbf {\bibinfo {volume}
  {105}},\ \bibinfo {pages} {044314} (\bibinfo {year} {2022})}\BibitemShut
  {NoStop}%
\bibitem [{\citenamefont {Liu}\ \emph {et~al.}(2024)\citenamefont {Liu},
  \citenamefont {Engel}, \citenamefont {Hinohara},\ and\ \citenamefont
  {Kortelainen}}]{Liu:2023xlv}%
  \BibitemOpen
  \bibfield  {author} {\bibinfo {author} {\bibfnamefont {Q.}~\bibnamefont
  {Liu}}, \bibinfo {author} {\bibfnamefont {J.}~\bibnamefont {Engel}}, \bibinfo
  {author} {\bibfnamefont {N.}~\bibnamefont {Hinohara}},\ and\ \bibinfo
  {author} {\bibfnamefont {M.}~\bibnamefont {Kortelainen}},\ }\href
  {https://doi.org/10.1103/PhysRevC.109.044308} {\bibfield  {journal} {\bibinfo
   {journal} {Phys. Rev. C}\ }\textbf {\bibinfo {volume} {109}},\ \bibinfo
  {pages} {044308} (\bibinfo {year} {2024})}\BibitemShut {NoStop}%
\bibitem [{\citenamefont {Washiyama}\ \emph {et~al.}(2021)\citenamefont
  {Washiyama}, \citenamefont {Hinohara},\ and\ \citenamefont
  {Nakatsukasa}}]{Washiyama:2020qfr}%
  \BibitemOpen
  \bibfield  {author} {\bibinfo {author} {\bibfnamefont {K.}~\bibnamefont
  {Washiyama}}, \bibinfo {author} {\bibfnamefont {N.}~\bibnamefont
  {Hinohara}},\ and\ \bibinfo {author} {\bibfnamefont {T.}~\bibnamefont
  {Nakatsukasa}},\ }\href {https://doi.org/10.1103/PhysRevC.103.014306}
  {\bibfield  {journal} {\bibinfo  {journal} {Phys. Rev. C}\ }\textbf {\bibinfo
  {volume} {103}},\ \bibinfo {pages} {014306} (\bibinfo {year}
  {2021})}\BibitemShut {NoStop}%
\bibitem [{\citenamefont {Bjelčić}\ and\ \citenamefont
  {Schunck}(2025)}]{BJELCIC2025109387}%
  \BibitemOpen
  \bibfield  {author} {\bibinfo {author} {\bibfnamefont {A.}~\bibnamefont
  {Bjelčić}}\ and\ \bibinfo {author} {\bibfnamefont {N.}~\bibnamefont
  {Schunck}},\ }\href
  {https://doi.org/https://doi.org/10.1016/j.cpc.2024.109387} {\bibfield
  {journal} {\bibinfo  {journal} {Computer Physics Communications}\ }\textbf
  {\bibinfo {volume} {306}},\ \bibinfo {pages} {109387} (\bibinfo {year}
  {2025})}\BibitemShut {NoStop}%
\bibitem [{\citenamefont {Vautherin}(1973)}]{Vautherin:1973zz}%
  \BibitemOpen
  \bibfield  {author} {\bibinfo {author} {\bibfnamefont {D.}~\bibnamefont
  {Vautherin}},\ }\href {https://doi.org/10.1103/PhysRevC.7.296} {\bibfield
  {journal} {\bibinfo  {journal} {Phys. Rev. C}\ }\textbf {\bibinfo {volume}
  {7}},\ \bibinfo {pages} {296} (\bibinfo {year} {1973})}\BibitemShut {NoStop}%
\bibitem [{\citenamefont {Bender}\ \emph {et~al.}(2003)\citenamefont {Bender},
  \citenamefont {Heenen},\ and\ \citenamefont {Reinhard}}]{Bender:2003jk}%
  \BibitemOpen
  \bibfield  {author} {\bibinfo {author} {\bibfnamefont {M.}~\bibnamefont
  {Bender}}, \bibinfo {author} {\bibfnamefont {P.-H.}\ \bibnamefont {Heenen}},\
  and\ \bibinfo {author} {\bibfnamefont {P.-G.}\ \bibnamefont {Reinhard}},\
  }\href {https://doi.org/10.1103/RevModPhys.75.121} {\bibfield  {journal}
  {\bibinfo  {journal} {Rev. Mod. Phys.}\ }\textbf {\bibinfo {volume} {75}},\
  \bibinfo {pages} {121} (\bibinfo {year} {2003})}\BibitemShut {NoStop}%
\bibitem [{\citenamefont {Satchler}(1964)}]{SATCHLER19641}%
  \BibitemOpen
  \bibfield  {author} {\bibinfo {author} {\bibfnamefont {G.}~\bibnamefont
  {Satchler}},\ }\href
  {https://doi.org/https://doi.org/10.1016/0029-5582(64)90124-5} {\bibfield
  {journal} {\bibinfo  {journal} {Nuclear Physics}\ }\textbf {\bibinfo {volume}
  {55}},\ \bibinfo {pages} {1} (\bibinfo {year} {1964})}\BibitemShut {NoStop}%
\bibitem [{\citenamefont {Maruhn}\ \emph {et~al.}(2014)\citenamefont {Maruhn},
  \citenamefont {Reinhard}, \citenamefont {Stevenson},\ and\ \citenamefont
  {Umar}}]{Maruhn:2013mpa}%
  \BibitemOpen
  \bibfield  {author} {\bibinfo {author} {\bibfnamefont {J.~A.}\ \bibnamefont
  {Maruhn}}, \bibinfo {author} {\bibfnamefont {P.~G.}\ \bibnamefont
  {Reinhard}}, \bibinfo {author} {\bibfnamefont {P.~D.}\ \bibnamefont
  {Stevenson}},\ and\ \bibinfo {author} {\bibfnamefont {A.~S.}\ \bibnamefont
  {Umar}},\ }\href {https://doi.org/10.1016/j.cpc.2014.04.008} {\bibfield
  {journal} {\bibinfo  {journal} {Comput. Phys. Commun.}\ }\textbf {\bibinfo
  {volume} {185}},\ \bibinfo {pages} {2195} (\bibinfo {year}
  {2014})}\BibitemShut {NoStop}%
\bibitem [{\citenamefont {Vesely}\ \emph {et~al.}(2009)\citenamefont {Vesely},
  \citenamefont {Kvasil}, \citenamefont {Nesterenko}, \citenamefont {Kleinig},
  \citenamefont {Reinhard},\ and\ \citenamefont {Ponomarev}}]{Vesely:2009eb}%
  \BibitemOpen
  \bibfield  {author} {\bibinfo {author} {\bibfnamefont {P.}~\bibnamefont
  {Vesely}}, \bibinfo {author} {\bibfnamefont {J.}~\bibnamefont {Kvasil}},
  \bibinfo {author} {\bibfnamefont {V.~O.}\ \bibnamefont {Nesterenko}},
  \bibinfo {author} {\bibfnamefont {W.}~\bibnamefont {Kleinig}}, \bibinfo
  {author} {\bibfnamefont {P.~G.}\ \bibnamefont {Reinhard}},\ and\ \bibinfo
  {author} {\bibfnamefont {V.~Y.}\ \bibnamefont {Ponomarev}},\ }\href
  {https://doi.org/10.1103/PhysRevC.80.031302} {\bibfield  {journal} {\bibinfo
  {journal} {Phys. Rev. C}\ }\textbf {\bibinfo {volume} {80}},\ \bibinfo
  {pages} {031302} (\bibinfo {year} {2009})}\BibitemShut {NoStop}%
\bibitem [{\citenamefont {Davies}\ and\ \citenamefont
  {Satchler}(1974)}]{Davies:1974ell}%
  \BibitemOpen
  \bibfield  {author} {\bibinfo {author} {\bibfnamefont {K.~T.~R.}\
  \bibnamefont {Davies}}\ and\ \bibinfo {author} {\bibfnamefont {G.~R.}\
  \bibnamefont {Satchler}},\ }\href
  {https://doi.org/10.1016/0375-9474(74)90582-X} {\bibfield  {journal}
  {\bibinfo  {journal} {Nucl. Phys. A}\ }\textbf {\bibinfo {volume} {222}},\
  \bibinfo {pages} {13} (\bibinfo {year} {1974})}\BibitemShut {NoStop}%
\bibitem [{\citenamefont {Sharp}\ and\ \citenamefont
  {Zamick}(1973)}]{Sharp1973130}%
  \BibitemOpen
  \bibfield  {author} {\bibinfo {author} {\bibfnamefont {R.}~\bibnamefont
  {Sharp}}\ and\ \bibinfo {author} {\bibfnamefont {L.}~\bibnamefont {Zamick}},\
  }\href {https://doi.org/https://doi.org/10.1016/0375-9474(73)90739-2}
  {\bibfield  {journal} {\bibinfo  {journal} {Nuclear Physics A}\ }\textbf
  {\bibinfo {volume} {208}},\ \bibinfo {pages} {130} (\bibinfo {year}
  {1973})}\BibitemShut {NoStop}%
\bibitem [{\citenamefont {{Barrera}}\ \emph {et~al.}(1985)\citenamefont
  {{Barrera}}, \citenamefont {{Estevez}},\ and\ \citenamefont
  {{Giraldo}}}]{Barrera1985EJPh}%
  \BibitemOpen
  \bibfield  {author} {\bibinfo {author} {\bibfnamefont {R.~G.}\ \bibnamefont
  {{Barrera}}}, \bibinfo {author} {\bibfnamefont {G.~A.}\ \bibnamefont
  {{Estevez}}},\ and\ \bibinfo {author} {\bibfnamefont {J.}~\bibnamefont
  {{Giraldo}}},\ }\href {https://doi.org/10.1088/0143-0807/6/4/014} {\bibfield
  {journal} {\bibinfo  {journal} {European Journal of Physics}\ }\textbf
  {\bibinfo {volume} {6}},\ \bibinfo {pages} {287} (\bibinfo {year}
  {1985})}\BibitemShut {NoStop}%
\bibitem [{\citenamefont {Johnson}\ \emph {et~al.}(1966)\citenamefont
  {Johnson}, \citenamefont {Owen},\ and\ \citenamefont
  {Satchler}}]{Johnson:1966zz}%
  \BibitemOpen
  \bibfield  {author} {\bibinfo {author} {\bibfnamefont {M.~B.}\ \bibnamefont
  {Johnson}}, \bibinfo {author} {\bibfnamefont {L.~W.}\ \bibnamefont {Owen}},\
  and\ \bibinfo {author} {\bibfnamefont {G.~R.}\ \bibnamefont {Satchler}},\
  }\href {https://doi.org/10.1103/PhysRev.142.748} {\bibfield  {journal}
  {\bibinfo  {journal} {Phys. Rev.}\ }\textbf {\bibinfo {volume} {142}},\
  \bibinfo {pages} {748} (\bibinfo {year} {1966})}\BibitemShut {NoStop}%
\bibitem [{\citenamefont {Kerveno}\ \emph {et~al.}(2021)\citenamefont
  {Kerveno}, \citenamefont {Dupuis}, \citenamefont {Bacquias}, \citenamefont
  {Belloni}, \citenamefont {Bernard}, \citenamefont {Borcea}, \citenamefont
  {Boromiza}, \citenamefont {Capote}, \citenamefont {De~Saint~Jean},
  \citenamefont {Dessagne}, \citenamefont {Droh\'e}, \citenamefont {Henning},
  \citenamefont {Hilaire}, \citenamefont {Kawano}, \citenamefont {Leconte},
  \citenamefont {Nankov}, \citenamefont {Negret}, \citenamefont {Nyman},
  \citenamefont {Olacel}, \citenamefont {Plompen}, \citenamefont {Romain},
  \citenamefont {Rouki}, \citenamefont {Rudolf}, \citenamefont {Stanoiu},\ and\
  \citenamefont {Wynants}}]{Kerveno2021}%
  \BibitemOpen
  \bibfield  {author} {\bibinfo {author} {\bibfnamefont {M.}~\bibnamefont
  {Kerveno}}, \bibinfo {author} {\bibfnamefont {M.}~\bibnamefont {Dupuis}},
  \bibinfo {author} {\bibfnamefont {A.}~\bibnamefont {Bacquias}}, \bibinfo
  {author} {\bibfnamefont {F.}~\bibnamefont {Belloni}}, \bibinfo {author}
  {\bibfnamefont {D.}~\bibnamefont {Bernard}}, \bibinfo {author} {\bibfnamefont
  {C.}~\bibnamefont {Borcea}}, \bibinfo {author} {\bibfnamefont
  {M.}~\bibnamefont {Boromiza}}, \bibinfo {author} {\bibfnamefont
  {R.}~\bibnamefont {Capote}}, \bibinfo {author} {\bibfnamefont
  {C.}~\bibnamefont {De~Saint~Jean}}, \bibinfo {author} {\bibfnamefont
  {P.}~\bibnamefont {Dessagne}}, \bibinfo {author} {\bibfnamefont {J.~C.}\
  \bibnamefont {Droh\'e}}, \bibinfo {author} {\bibfnamefont {G.}~\bibnamefont
  {Henning}}, \bibinfo {author} {\bibfnamefont {S.}~\bibnamefont {Hilaire}},
  \bibinfo {author} {\bibfnamefont {T.}~\bibnamefont {Kawano}}, \bibinfo
  {author} {\bibfnamefont {P.}~\bibnamefont {Leconte}}, \bibinfo {author}
  {\bibfnamefont {N.}~\bibnamefont {Nankov}}, \bibinfo {author} {\bibfnamefont
  {A.}~\bibnamefont {Negret}}, \bibinfo {author} {\bibfnamefont
  {M.}~\bibnamefont {Nyman}}, \bibinfo {author} {\bibfnamefont
  {A.}~\bibnamefont {Olacel}}, \bibinfo {author} {\bibfnamefont {A.~J.~M.}\
  \bibnamefont {Plompen}}, \bibinfo {author} {\bibfnamefont {P.}~\bibnamefont
  {Romain}}, \bibinfo {author} {\bibfnamefont {C.}~\bibnamefont {Rouki}},
  \bibinfo {author} {\bibfnamefont {G.}~\bibnamefont {Rudolf}}, \bibinfo
  {author} {\bibfnamefont {M.}~\bibnamefont {Stanoiu}},\ and\ \bibinfo {author}
  {\bibfnamefont {R.}~\bibnamefont {Wynants}},\ }\href
  {https://doi.org/10.1103/PhysRevC.104.044605} {\bibfield  {journal} {\bibinfo
   {journal} {Phys. Rev. C}\ }\textbf {\bibinfo {volume} {104}},\ \bibinfo
  {pages} {044605} (\bibinfo {year} {2021})}\BibitemShut {NoStop}%
\bibitem [{\citenamefont {Martin}(2007)}]{Martin20071583}%
  \BibitemOpen
  \bibfield  {author} {\bibinfo {author} {\bibfnamefont {M.}~\bibnamefont
  {Martin}},\ }\href
  {https://doi.org/https://doi.org/10.1016/j.nds.2007.07.001} {\bibfield
  {journal} {\bibinfo  {journal} {Nuclear Data Sheets}\ }\textbf {\bibinfo
  {volume} {108}},\ \bibinfo {pages} {1583} (\bibinfo {year}
  {2007})}\BibitemShut {NoStop}%
\bibitem [{\citenamefont {Kinney}\ and\ \citenamefont
  {Perey}(1974)}]{Kinney_Perey_1974}%
  \BibitemOpen
  \bibfield  {author} {\bibinfo {author} {\bibfnamefont {W.}~\bibnamefont
  {Kinney}}\ and\ \bibinfo {author} {\bibfnamefont {F.}~\bibnamefont {Perey}},\
  }\href {https://doi.org/10.2172/4276980} {\bibfield  {journal} {\bibinfo
  {journal} {Oak Ridge National Laboratory Reports}\ ,\ \bibinfo {pages}
  {Tech.Rep.4909}} (\bibinfo {year} {1974})}\BibitemShut {NoStop}%
\bibitem [{\citenamefont {Bainum}\ \emph {et~al.}(1977)\citenamefont {Bainum},
  \citenamefont {Finlay}, \citenamefont {Rapaport}, \citenamefont {Carlson},\
  and\ \citenamefont {Love}}]{Bainum1977}%
  \BibitemOpen
  \bibfield  {author} {\bibinfo {author} {\bibfnamefont {D.~E.}\ \bibnamefont
  {Bainum}}, \bibinfo {author} {\bibfnamefont {R.~W.}\ \bibnamefont {Finlay}},
  \bibinfo {author} {\bibfnamefont {J.}~\bibnamefont {Rapaport}}, \bibinfo
  {author} {\bibfnamefont {J.~D.}\ \bibnamefont {Carlson}},\ and\ \bibinfo
  {author} {\bibfnamefont {W.~G.}\ \bibnamefont {Love}},\ }\href
  {https://doi.org/10.1103/PhysRevC.16.1377} {\bibfield  {journal} {\bibinfo
  {journal} {Phys. Rev. C}\ }\textbf {\bibinfo {volume} {16}},\ \bibinfo
  {pages} {1377} (\bibinfo {year} {1977})}\BibitemShut {NoStop}%
\bibitem [{\citenamefont {Haouat}\ \emph
  {et~al.}(1982{\natexlab{a}})\citenamefont {Haouat}, \citenamefont
  {Bersillon}, \citenamefont {Lachkar}, \citenamefont {Patin},\ and\
  \citenamefont {Lagrange}}]{Haouat1982r2200}%
  \BibitemOpen
  \bibfield  {author} {\bibinfo {author} {\bibfnamefont {G.}~\bibnamefont
  {Haouat}}, \bibinfo {author} {\bibfnamefont {O.}~\bibnamefont {Bersillon}},
  \bibinfo {author} {\bibfnamefont {J.}~\bibnamefont {Lachkar}}, \bibinfo
  {author} {\bibfnamefont {Y.}~\bibnamefont {Patin}},\ and\ \bibinfo {author}
  {\bibfnamefont {C.}~\bibnamefont {Lagrange}},\ }\href@noop {} {\bibfield
  {journal} {\bibinfo  {journal} {CEA report No. CEA-N-2200 (France)}\ }
  (\bibinfo {year} {1982}{\natexlab{a}})}\BibitemShut {NoStop}%
\bibitem [{\citenamefont {Haouat}\ \emph
  {et~al.}(1982{\natexlab{b}})\citenamefont {Haouat}, \citenamefont {Dietrich},
  \citenamefont {Patin}, \citenamefont {De~Swiniarski}, \citenamefont {Humeau},
  \citenamefont {Chardine}, \citenamefont {P.~Lochard},\ and\ \citenamefont
  {Seguin}}]{Haouat1982r2284}%
  \BibitemOpen
  \bibfield  {author} {\bibinfo {author} {\bibfnamefont {G.}~\bibnamefont
  {Haouat}}, \bibinfo {author} {\bibfnamefont {F.}~\bibnamefont {Dietrich}},
  \bibinfo {author} {\bibfnamefont {Y.}~\bibnamefont {Patin}}, \bibinfo
  {author} {\bibfnamefont {R.}~\bibnamefont {De~Swiniarski}}, \bibinfo {author}
  {\bibfnamefont {C.}~\bibnamefont {Humeau}}, \bibinfo {author} {\bibfnamefont
  {J.}~\bibnamefont {Chardine}}, \bibinfo {author} {\bibfnamefont
  {J.}~\bibnamefont {P.~Lochard}},\ and\ \bibinfo {author} {\bibfnamefont
  {S.}~\bibnamefont {Seguin}},\ }\href@noop {} {\bibfield  {journal} {\bibinfo
  {journal} {CEA report No. CEA-N-2284 (France)}\ } (\bibinfo {year}
  {1982}{\natexlab{b}})}\BibitemShut {NoStop}%
\bibitem [{\citenamefont {Bonneau}\ \emph {et~al.}(2007)\citenamefont
  {Bonneau}, \citenamefont {Kawano}, \citenamefont {Watanabe},\ and\
  \citenamefont {Chiba}}]{Bonneau:2007dc}%
  \BibitemOpen
  \bibfield  {author} {\bibinfo {author} {\bibfnamefont {L.}~\bibnamefont
  {Bonneau}}, \bibinfo {author} {\bibfnamefont {T.}~\bibnamefont {Kawano}},
  \bibinfo {author} {\bibfnamefont {T.}~\bibnamefont {Watanabe}},\ and\
  \bibinfo {author} {\bibfnamefont {S.}~\bibnamefont {Chiba}},\ }\href
  {https://doi.org/10.1103/PhysRevC.75.054618} {\bibfield  {journal} {\bibinfo
  {journal} {Phys. Rev. C}\ }\textbf {\bibinfo {volume} {75}},\ \bibinfo
  {pages} {054618} (\bibinfo {year} {2007})}\BibitemShut {NoStop}%
\bibitem [{\citenamefont {Chabanat}\ \emph {et~al.}(1998)\citenamefont
  {Chabanat}, \citenamefont {Bonche}, \citenamefont {Haensel}, \citenamefont
  {Meyer},\ and\ \citenamefont {Schaeffer}}]{Chabanat:1997un}%
  \BibitemOpen
  \bibfield  {author} {\bibinfo {author} {\bibfnamefont {E.}~\bibnamefont
  {Chabanat}}, \bibinfo {author} {\bibfnamefont {P.}~\bibnamefont {Bonche}},
  \bibinfo {author} {\bibfnamefont {P.}~\bibnamefont {Haensel}}, \bibinfo
  {author} {\bibfnamefont {J.}~\bibnamefont {Meyer}},\ and\ \bibinfo {author}
  {\bibfnamefont {R.}~\bibnamefont {Schaeffer}},\ }\href
  {https://doi.org/10.1016/S0375-9474(98)00180-8} {\bibfield  {journal}
  {\bibinfo  {journal} {Nucl. Phys. A}\ }\textbf {\bibinfo {volume} {635}},\
  \bibinfo {pages} {231} (\bibinfo {year} {1998})}\BibitemShut {NoStop}%
\bibitem [{\citenamefont {Kawano}(2021)}]{Kawano2021}%
  \BibitemOpen
  \bibfield  {author} {\bibinfo {author} {\bibfnamefont {T.}~\bibnamefont
  {Kawano}},\ }\href {https://doi.org/10.1140/epja/s10050-020-00311-9}
  {\bibfield  {journal} {\bibinfo  {journal} {European Physical Journal A}\
  }\textbf {\bibinfo {volume} {57}},\ \bibinfo {pages} {16} (\bibinfo {year}
  {2021})}\BibitemShut {NoStop}%
\bibitem [{\citenamefont {Kunieda}\ \emph {et~al.}(2007)\citenamefont
  {Kunieda}, \citenamefont {Chiba}, \citenamefont {Shibata}, \citenamefont
  {Ichihara},\ and\ \citenamefont {Sukhovitsk{\~{i}}}}]{Kunieda2007}%
  \BibitemOpen
  \bibfield  {author} {\bibinfo {author} {\bibfnamefont {S.}~\bibnamefont
  {Kunieda}}, \bibinfo {author} {\bibfnamefont {S.}~\bibnamefont {Chiba}},
  \bibinfo {author} {\bibfnamefont {K.}~\bibnamefont {Shibata}}, \bibinfo
  {author} {\bibfnamefont {A.}~\bibnamefont {Ichihara}},\ and\ \bibinfo
  {author} {\bibfnamefont {E.~S.}\ \bibnamefont {Sukhovitsk{\~{i}}}},\ }\href
  {https://doi.org/10.1080/18811248.2007.9711321} {\bibfield  {journal}
  {\bibinfo  {journal} {Journal of Nuclear Science and Technology}\ }\textbf
  {\bibinfo {volume} {44}},\ \bibinfo {pages} {838} (\bibinfo {year}
  {2007})}\BibitemShut {NoStop}%
\bibitem [{\citenamefont {Perey}\ and\ \citenamefont
  {Saxon}(1964)}]{Perey1964}%
  \BibitemOpen
  \bibfield  {author} {\bibinfo {author} {\bibfnamefont {F.}~\bibnamefont
  {Perey}}\ and\ \bibinfo {author} {\bibfnamefont {D.}~\bibnamefont {Saxon}},\
  }\href {https://doi.org/https://doi.org/10.1016/0031-9163(64)90597-9}
  {\bibfield  {journal} {\bibinfo  {journal} {Physics Letters}\ }\textbf
  {\bibinfo {volume} {10}},\ \bibinfo {pages} {107} (\bibinfo {year}
  {1964})}\BibitemShut {NoStop}%
\bibitem [{\citenamefont {Finlay}\ \emph {et~al.}(1984)\citenamefont {Finlay},
  \citenamefont {Annand}, \citenamefont {Cheema}, \citenamefont {Rapaport},\
  and\ \citenamefont {Dietrich}}]{Finlay:1984prc}%
  \BibitemOpen
  \bibfield  {author} {\bibinfo {author} {\bibfnamefont {R.~W.}\ \bibnamefont
  {Finlay}}, \bibinfo {author} {\bibfnamefont {J.~R.~M.}\ \bibnamefont
  {Annand}}, \bibinfo {author} {\bibfnamefont {T.~S.}\ \bibnamefont {Cheema}},
  \bibinfo {author} {\bibfnamefont {J.}~\bibnamefont {Rapaport}},\ and\
  \bibinfo {author} {\bibfnamefont {F.~S.}\ \bibnamefont {Dietrich}},\ }\href
  {https://doi.org/10.1103/PhysRevC.30.796} {\bibfield  {journal} {\bibinfo
  {journal} {Phys. Rev. C}\ }\textbf {\bibinfo {volume} {30}},\ \bibinfo
  {pages} {796} (\bibinfo {year} {1984})}\BibitemShut {NoStop}%
\bibitem [{\citenamefont {{Takahashi}}\ \emph {et~al.}(1988)\citenamefont
  {{Takahashi}}, \citenamefont {{Ichimura}}, \citenamefont {{Sasaki}},\ and\
  \citenamefont {{Sugimoto}}}]{Takahashi01031988}%
  \BibitemOpen
  \bibfield  {author} {\bibinfo {author} {\bibfnamefont {A.}~\bibnamefont
  {{Takahashi}}}, \bibinfo {author} {\bibfnamefont {E.}~\bibnamefont
  {{Ichimura}}}, \bibinfo {author} {\bibfnamefont {Y.}~\bibnamefont
  {{Sasaki}}},\ and\ \bibinfo {author} {\bibfnamefont {H.}~\bibnamefont
  {{Sugimoto}}},\ }\href {https://doi.org/10.1080/18811248.1988.9733582}
  {\bibfield  {journal} {\bibinfo  {journal} {Journal of Nuclear Science and
  Technology}\ }\textbf {\bibinfo {volume} {25}},\ \bibinfo {pages} {215}
  (\bibinfo {year} {1988})}\BibitemShut {NoStop}%
\bibitem [{\citenamefont {Harakeh}\ and\ \citenamefont
  {Woude}(2001)}]{harakeh2001giant}%
  \BibitemOpen
  \bibfield  {author} {\bibinfo {author} {\bibfnamefont {M.}~\bibnamefont
  {Harakeh}}\ and\ \bibinfo {author} {\bibfnamefont {A.}~\bibnamefont
  {Woude}},\ }\href@noop {} {\emph {\bibinfo {title} {Giant Resonances:
  Fundamental High-frequency Modes of Nuclear Excitation}}}\ (\bibinfo
  {publisher} {Oxford University Press},\ \bibinfo {year} {2001})\BibitemShut
  {NoStop}%
\bibitem [{\citenamefont {Ring}\ and\ \citenamefont
  {Speth}(1973)}]{RING1973477}%
  \BibitemOpen
  \bibfield  {author} {\bibinfo {author} {\bibfnamefont {P.}~\bibnamefont
  {Ring}}\ and\ \bibinfo {author} {\bibfnamefont {J.}~\bibnamefont {Speth}},\
  }\href {https://doi.org/https://doi.org/10.1016/0370-2693(73)90001-4}
  {\bibfield  {journal} {\bibinfo  {journal} {Physics Letters B}\ }\textbf
  {\bibinfo {volume} {44}},\ \bibinfo {pages} {477} (\bibinfo {year}
  {1973})}\BibitemShut {NoStop}%
\bibitem [{\citenamefont {Roca-Maza}\ \emph {et~al.}(2013)\citenamefont
  {Roca-Maza}, \citenamefont {Brenna}, \citenamefont {Agrawal}, \citenamefont
  {Bortignon}, \citenamefont {Col\`o}, \citenamefont {Cao}, \citenamefont
  {Paar},\ and\ \citenamefont {Vretenar}}]{Roca-MazaPRC2013}%
  \BibitemOpen
  \bibfield  {author} {\bibinfo {author} {\bibfnamefont {X.}~\bibnamefont
  {Roca-Maza}}, \bibinfo {author} {\bibfnamefont {M.}~\bibnamefont {Brenna}},
  \bibinfo {author} {\bibfnamefont {B.~K.}\ \bibnamefont {Agrawal}}, \bibinfo
  {author} {\bibfnamefont {P.~F.}\ \bibnamefont {Bortignon}}, \bibinfo {author}
  {\bibfnamefont {G.}~\bibnamefont {Col\`o}}, \bibinfo {author} {\bibfnamefont
  {L.-G.}\ \bibnamefont {Cao}}, \bibinfo {author} {\bibfnamefont
  {N.}~\bibnamefont {Paar}},\ and\ \bibinfo {author} {\bibfnamefont
  {D.}~\bibnamefont {Vretenar}},\ }\href
  {https://doi.org/10.1103/PhysRevC.87.034301} {\bibfield  {journal} {\bibinfo
  {journal} {Phys. Rev. C}\ }\textbf {\bibinfo {volume} {87}},\ \bibinfo
  {pages} {034301} (\bibinfo {year} {2013})}\BibitemShut {NoStop}%
\bibitem [{\citenamefont {Gilbert}\ and\ \citenamefont
  {Cameron}(1965)}]{Gilbert:1965vs}%
  \BibitemOpen
  \bibfield  {author} {\bibinfo {author} {\bibfnamefont {A.}~\bibnamefont
  {Gilbert}}\ and\ \bibinfo {author} {\bibfnamefont {A.~G.~W.}\ \bibnamefont
  {Cameron}},\ }\href {https://doi.org/10.1139/p65-139} {\bibfield  {journal}
  {\bibinfo  {journal} {Can. J. Phys.}\ }\textbf {\bibinfo {volume} {43}},\
  \bibinfo {pages} {1446} (\bibinfo {year} {1965})}\BibitemShut {NoStop}%
\bibitem [{\citenamefont {M\"{o}ller}\ \emph {et~al.}(1995)\citenamefont
  {M\"{o}ller}, \citenamefont {Nix}, \citenamefont {Myer},\ and\ \citenamefont
  {Swiatecki}}]{Moller1995}%
  \BibitemOpen
  \bibfield  {author} {\bibinfo {author} {\bibfnamefont {P.}~\bibnamefont
  {M\"{o}ller}}, \bibinfo {author} {\bibfnamefont {J.~R.}\ \bibnamefont {Nix}},
  \bibinfo {author} {\bibfnamefont {W.~D.}\ \bibnamefont {Myer}},\ and\
  \bibinfo {author} {\bibfnamefont {W.~J.}\ \bibnamefont {Swiatecki}},\
  }\bibfield  {title} {\bibinfo {title} {Nuclear ground-state masses and
  deformations},\ }\href {https://doi.org/10.1006/adnd.1995.1002} {\bibfield
  {journal} {\bibinfo  {journal} {Atomic Data and Nuclear Data Tables}\
  }\textbf {\bibinfo {volume} {59}},\ \bibinfo {pages} {185 } (\bibinfo {year}
  {1995})}\BibitemShut {NoStop}%
\bibitem [{\citenamefont {M\"{o}ller}\ \emph {et~al.}(2016)\citenamefont
  {M\"{o}ller}, \citenamefont {Sierk}, \citenamefont {Ichikawa},\ and\
  \citenamefont {Sagawa}}]{Moller2016}%
  \BibitemOpen
  \bibfield  {author} {\bibinfo {author} {\bibfnamefont {P.}~\bibnamefont
  {M\"{o}ller}}, \bibinfo {author} {\bibfnamefont {A.~J.}\ \bibnamefont
  {Sierk}}, \bibinfo {author} {\bibfnamefont {T.}~\bibnamefont {Ichikawa}},\
  and\ \bibinfo {author} {\bibfnamefont {H.}~\bibnamefont {Sagawa}},\
  }\bibfield  {title} {\bibinfo {title} {Nuclear ground-state masses and
  deformations: Frdm(2012)},\ }\href
  {https://doi.org/10.1016/j.adt.2015.10.002} {\bibfield  {journal} {\bibinfo
  {journal} {Atomic Data and Nuclear Data Tables}\ }\textbf {\bibinfo {volume}
  {109 -- 110}},\ \bibinfo {pages} {1 } (\bibinfo {year} {2016})}\BibitemShut
  {NoStop}%
\bibitem [{\citenamefont {Gruppelaar}(1983)}]{Gruppelaar1983}%
  \BibitemOpen
  \bibfield  {author} {\bibinfo {author} {\bibfnamefont {H.}~\bibnamefont
  {Gruppelaar}},\ }\href@noop {} {\emph {\bibinfo {title} {Level density in
  unified preequilibrium and equilibrium models}}},\ \bibinfo {type} {Tech.
  Rep.}\ (\bibinfo  {institution} {Brookhaven National Laboratory},\ \bibinfo
  {year} {1983})\BibitemShut {NoStop}%
\end{thebibliography}%


\end{document}